\documentclass{sn-jnl}
\usepackage{subfigure}
\usepackage{url}
\usepackage{booktabs,tabularx}
\usepackage{amsfonts}
\usepackage{amssymb}
\usepackage{amsmath}
\usepackage{mathrsfs}
\usepackage{graphicx}
\usepackage{commath}
\usepackage{amsthm}
\usepackage{multirow}
\usepackage{extarrows}
\usepackage{flushend}
\DeclareMathAlphabet{\mathpzc}{OT1}{pzc}{m}{it}
\newtheorem{defn}{Definition}[section]
\newtheorem{theorem}{Theorem}[section]
\newtheorem{lemma}{Lemma}[section]
\begin{document}

\title[Efficient Subgraph Isomorphism Finding]{Efficient Subgraph Isomorphism Finding in Large Graphs
using Eccentricity and Limiting Recursive Calls}
\author[1]{\fnm{Zubair Ali} \sur{Ansari}}\email{zuberaliansari@gmail.com}

\author[2]{\fnm{Muhammad} \sur{Abulaish}}\email{abulaish@sau.ac.in}
\author[3]{\fnm{Irfan Rashid Thoker }\email{zabooirfan@gmail.com}}
\equalcont{These authors contributed equally to this work.}

\author[4]{\fnm{Jahiruddin}}\email{jahir.jmi@gmail.com}
\equalcont{These authors contributed equally to this work.}
\affil[1]{\orgdiv{Department of Computer Science and Engineering}, \orgname{Gheethanjali College of Engineering and Technology}, \city{Telangana}, \country{India}}

\affil[2]{\orgdiv{Department of Computer Science}, \orgname{South Asian University}, \orgaddress{\city{New Delhi}, \country{India}}}
\affil[3]{\orgdiv{Department of Education}, \orgname{Govt. of Jammu Kashmir}, \city{Jammu and Kashmir}, \country{India}}

\affil[4]{\orgdiv{Department of Computer Science}, \orgname{Jamia Millia Islamia (A Central University)}, \city{New Delhi}, \country{India}}

\maketitle
\section*{Abstract}
\textbf{Purpose} – The subgraph isomorphism finding problem is a well-studied problem in the field of computer science and graph theory, and it aims to enumerate all instances of a query graph in the respective data graph. Most of the existing algorithms, including RI,  QuickSI, and TurboISO minimize the computing cost of subgraph isomorphism finding problem through reducing the number and size of candidate regions in the data graph that possibly contains some instances of the query graph. The purpose of this paper is to propose an efficient method, SubISO, to find subgraph isomorphisms using an objective function, which exploits some isomorphic invariants and eccentricity of the query graph's vertices.\\
\textbf{Design/Methodology/Approach} – The proposed objective function is used to determine pivot vertex, which minimizes both number and size of the candidate regions in the data graph. SubISO also limits the maximum recursive calls of the generic SubgraphSearch() function to deal with straggler queries for which most of the existing algorithms show exponential behaviour.\\
\textbf{Findings} – The proposed approach is evaluated over three benchmark datasets. It is also compared with three well-known subgraph isomorphism finding algorithms in terms of execution time, number of identified embeddings, and ability to deal with the straggler queries, and it performs significantly better.\\
\textbf{Research Implications} – An attempt is made to develop scalable and effective algorithms for the analysis of huge graphs obtained from social networks, biological networks, and Resource Description Framework (RDF) data. An effective method for finding subgraph isomorphisms is introduced: SubISO. The performance of the suggested algorithm is assessed using the most advanced methods currently available. The suggested algorithm performed better than the alternative methods.\\
\textbf{Originality/Value} – This paper fulfils an identified need to study subgraph isomorphism finding problem.\\
\textbf{Keywords}: Subgraph Isomorphism, Graph Search, Eccentricity, Candidate Region, Large Graph, Embedding, Straggler Query.\\
\textbf{Paper type}: Research Paper.

\section{Introduction}\label{sec:Introduction}
Structural information extraction from large graphs in a reasonable time plays a vital role in solving many real-world problems \cite{conte2004thirty}, \cite{foggia2014graph}. A graph is a well-known mathematical formalization to model connected data and the underlying structural information that can be analyzed for varied purposes, including community detection, information diffusion, and so on. Subgraph isomorphism finding is a well-known NP-complete problem in network science which mainly aims to identify the regions (\textit{aka} subgraphs) of a data graph that are isomorphic images of the query graph \cite{gary1979computers}. In other words, the subgraph isomorphism finding problem aims to enumerate all isomorphic images of a query graph in the respective data graph. It becomes a central problem when we need to query a large data graph containing millions of vertices and edges. Though many researchers have proposed efficient algorithms like $\mathrm{Turbo_{ISO}}$ \cite{han2013turbo} and \texttt{QuickSI} \cite{shang2008taming} to deal with the subgraph isomorphism finding problem, the issues like scalability, time complexity, dependency on the nature of the query graph and data graph are still elusive.
\subsection{Pivot Node Selection Problem and Straggler Queries} \label{PNS}
In order to find all subgraph isomorphisms of a given query graph into a data graph, we need to explore all possible candidate regions of the data graph. For a given query graph, the candidate region is a portion of the data graph that may have some isomorphic images of the query graph. Therefore, for efficient subgraph isomorphism finding, we need to minimize both the number and size of the candidate regions. To this end, the selection of a pivot vertex (a vertex to start the matching process) of the query graph plays an important role and researchers have proposed different rank functions to determine the pivot vertex. For example, \texttt{STwig} algorithm \cite{sun2012efficient} used the rank function $Rank(u)=freq(G_d,l(u))/\Delta(u)$, where $freq(G_d,l(u))$ is the frequency of label $l(u)$ in data graph $G_d$ and $\Delta(u)$ is the degree of vertex $u$ in query graph $G_q$. It gives higher rank to a vertex of an infrequent label with a higher degree. $\mathrm{Turbo_{ISO}}$ \cite{han2013turbo} also used this rank function to choose pivot vertex of the query graph. On the other hand, \texttt{CFL-Match} \cite{bi2016efficient} used the rank function $\underset{u}{\arg\min} (|C(u)|/\Delta(u))$, where $C(u)$ is the candidate set of query vertex $u$ in data graph $G_d$ and $\Delta(u)$ is the degree of vertex $u$ in query graph $G_q$. This rank function favour a vertex of minimum candidate sets with a higher degree. However, none of the above rank functions considers minimizing the size of candidate regions. 

A study in \cite{han2013turbo} revealed that generally subgraph isomorphism methods show exponential behavior for some combination of query and data graphs because subgraph isomorphism is a combinatorial problem demonstrated to be NP (nondeterministic polynomial time), mainly due to symmetries in the structure of a graph. Sometimes, matching order also causes exponential behavior for some subgraph isomorphism methods. Recently, authors in \cite{katsarou2018improving} investigated the exponential behavior of five existing subgraph isomorphism finding algorithms -- \texttt{GraphQL}, \texttt{SPath}, \texttt{QuickSI}, \texttt{$\mathrm{Turbo_{ISO}}$}, and \texttt{$\mathrm{Boost_{ISO}}$} on some combinations of query and data graphs. They named such query as \textit{straggler query} and defined as follows:

\begin{defn}{\texttt{Straggler query \cite{katsarou2018improving}:}} 
``Queries with execution times many orders of magnitude worse compared to the majority of them."
\end{defn}       

They reported that all existing methods have some straggler queries, and they are generally algorithm specific. They also proposed an alternative solution to deal with such queries that can be found in \cite{katsarou2018improving}.

\subsection{Our Contributions} \label{OC}
Our main contribution lies in the development a subgraph isomorphism finding method, $\mathrm{SubISO}$, which uses a novel objective function for determining pivot vertex that minimizes both number and size of the candidate regions in data graph. To this end, $\mathrm{SubISO}$ exploits some isomorphic invariants and eccentricity of the query graph's vertices. On analysis, we found that the generic \texttt{SubgraphSearch()} function uses an extensive number of recursive calls, and hence execution time raises exponentially in case of straggler queries. Therefore, we have introduced a new parameter, $\eta$, for restricting the maximum recursive calls of the generic \texttt{SubgraphSearch()} function to print at most $k$ number of solutions of the subgraph isomorphism finding problem efficiently. It also facilitates to deal with straggler queries for which most of the existing algorithms show exponential behavior.

In short, the major contributions of this work can be summarized as follows:
\begin{itemize}
	\item Introducing a novel objective function based on the eccentricity and some isomorphic invariants of the query graph vertices to minimize both number and size of candidate regions in data graph.
	\item Designing and implementing a new algorithm, $\mathrm{SubISO}$, to find at most $k$ solutions of the subgraph isomorphism finding problem.  
	\item Introducing a new parameter, $\eta$, for restricting the maximum recursive calls of the generic \texttt{SubgraphSearch()} function to handle straggler queries for which most of the existing algorithms show exponential behavior.
	\item Introducing a matching order, based on the degree of the nodes in query graph and number of matches in the candidate region of the data graph, to minimize the number of backtracking steps.
\end{itemize}

It may be noted that this work is a substantial extension of our initial study, which has been published in a conference proceedings \cite{abulaish2019subiso}. The substantial extension mainly includes (i) a formal description and mathematical proofs of the proposed objective function for pivot node selection, (ii) algorithm and executable codes of the proposed \texttt{$\mathrm{SubISO}$} algorithm for subgraph isomorphism finding, (iii) generation of query sets for three benchmark data graphs, (iv) extended experiments over additional datasets and comparison with some state-of-the-art methods, (v) a mechanism to deal with straggler queries, and (v) a mathematical proof to establish the \textit{soundness} and \textit{completeness} properties of the \texttt{$\mathrm{SubISO}$} algorithm.

The remaining portion of this paper is organized as follows. Section \ref{RelatedWorks} presents a review of the existing works on subgraph isomorphism finding problem. Section \ref{sec:BACKGROUND} presents some preliminaries including definitions, problem statement, and lemmas. Section \ref{sec:proposedmethod} presents the procedural details and pseudo codes of the proposed \texttt{$\mathrm{SubISO}$} algorithm. Section \ref{sec:Experiments} presents the experimental setup and evaluation results. It also presents a comparative analysis of \texttt{$\mathrm{SubISO}$} with some of the state-of-the-art methods. Section \ref{sec:Disscussion} presents a detailed discussion on limiting recursive call to deal with straggler queries. It also presents a mathematical proof to establish the soundness and completeness properties of the $\mathrm{SubISO}$ algorithm. Finally, section \ref{sec:ConclusionAndFutureWork} concludes the paper and provides some future directions of research.

\section{Related Works} \label{RelatedWorks}
This section presents a review of the existing works on subgraph isomorphism finding problem. 
Ullmann introduced the first algorithm to address the problem subgraph isomorphism finding in \cite{ullmann1976algorithm}. Subsequently, a large number of researchers proposed different solutions to solve the subgraph isomorphism finding problem. Some of the existing approaches are \texttt{VF2} \cite{cordella2004sub}, \texttt{GADDI} \cite{zhang2009gaddi}, \texttt{GraphQL} \cite{he2008graphs}, \texttt{SPath} \cite{zhao2010graph}, \texttt{STW} \cite{sun2012efficient}, \texttt{CFL-Match} \cite{bi2016efficient}, \texttt{$\mathrm{Sum_{ISO}}$} \cite{nabti2017querying}, \texttt{QuickSI} \cite{shang2008taming}, \texttt{$\mathrm{Turbo_{ISO}}$} \cite{han2013turbo}, and \texttt{BoostIso} \cite{ren2015exploiting}. 

\texttt{VF2} used a set of feasibility rules and suitable data structures during the matching process to reduce computational complexity. \texttt{QuickSI} enhanced the filtering and verification framework. They introduced QI-Sequence, encoded search order, and topological information of query graph to reduce verification cost. \texttt{GADDI} is an index based graph search algorithm, specially designed for biological networks. \texttt{GraphQL} utilized neighborhood signature based pruning  and pseudo subgraph isomorphism test to reduce the size of the candidate set. \texttt{SPath} introduced path-at-a-time, a new way to process a query graph using the shortest path based indexing to shrink search space. \texttt{STW} relied on Trinity memory cloud for basic graph exploration mechanism and does not use any graph indexing technique. \texttt{$\mathrm{Turbo_{ISO}}$} used two new concepts -- \textit{candidate region exploration}, and \textit{combine and permute strategy}. It used candidate region exploration to find a promising subgraph of the data graph that may have some instances of the query graph, whereas \textit{combine and permute strategy} takes advantage of neighborhood equivalence class to avoid blindly permutating all mappings of the query graph. \texttt{CFL-Match} considered the problem of the redundant Cartesian product by dissimilar vertices and proposed a new framework to decompose query graph into three sub-structures, namely \textit{core}, \textit{forest}, and \textit{leaves}. \texttt{$\mathrm{Sum_{ISO}}$} used graph compression with modular decomposition to find subgraph isomorphisms. \texttt{BoostIso} \cite{ren2015exploiting} uses \textit{syntactic containment, syntactic equivalence, query-dependent containment}, and \textit{query-dependent equivalence} to lower duplicate computation. They show these concepts could help to speed up subgraph isomorphism algorithms.  

An extensive comparative analysis of six subgraph isomorphism finding algorithms, namely \texttt{Ullmann},  \texttt{VF2}, \texttt{GADDI}, \texttt{QuickSI}, \texttt{SPath}, and  \texttt{GraphQL} was reported in \cite{lee2012depth}. They found that \texttt{VF2}, \texttt{QuickSI}, and \texttt{GADDI} are unable to find any embeddings of the query graph in tree-like structured data graph and show exponential behavior due to the join order selection. Authors in \cite{katsarou2018improving} analyzed the exponential behavior of  \texttt{GraphQL}, \texttt{sPath}, \texttt{QuickSI}, \texttt{$\mathrm{Turbo_{ISO}}$}, and \texttt{$\mathrm{Boost_{ISO}}$} on some query sets (straggler queries), and claimed that isomorphic counterpart of straggler queries could work well. The problem of \textit{output crisis} in subgraph matching was addressed in \cite{qiao2017subgraph}. In this paper, the authors proposed a \textit{vertex-cover based compression} technique to compress the output. Authors in \cite{bonnici2016variable} investigated the impact of vertex matching order on the running time of subgraph isomorphism finding algorithms in biological networks. Several improvements of the Ullmann's algorithm to reduce both time and space complexities have been discussed in \cite{vcibej2015improvements}. The enhanced Ullmann's algorithm are compared with \texttt{FS} \cite{ullmann2011bit}, \texttt{LAD} \cite{solnon2010alldifferent}, and \texttt{VF2} algorithms. Similarly, many enhanced versions of \texttt{VF2} viz. \texttt{VF2++} \cite{juttner2018vf2++}, \texttt{VF3-Light} \cite{carletti2019vf3}, and \texttt{VF3} \cite{carletti2017challenging} have also been proposed in literature. \texttt{VF3} has been proposed to find induced subgraph isomorphisms in extensive and very dense graphs, and its enhancement to solve the non-induced case of subgraph isomorphism is still under development \footnote{https://github.com/MiviaLab/vf3lib}. But, \texttt{VF3} is not compared with any other graph index-based subgraph isomorphism algorithm like \texttt{GraphQL}, \texttt{QuickSI}, \texttt{GADDI}, \texttt{SPath}, or $\mathrm{Turbo_{ISO}}$. \texttt{VF3-Light} is a simplified form of the \texttt{VF3}, and it is more effective on some types of graphs.

In \cite{bonnici2013subgraph}, the authors proposed \texttt{RI}, which generates a search strategy based on the pattern graph's topology, and it does not apply any computationally expensive pruning or inference rules during the matching process.In \cite{jian2023suff}, the authors  introduced SUFF, a versatile structure filtering framework for subgraph matching in graph theory. It accelerates existing approaches by creating filters based on past query results, significantly reducing search space and achieving up to 15X speedup with minimal overheads. The study in \cite{han2019efficient} addressed the problem of subgraph isomorphism using a spanning tree instead of a query graph, such as non-optimal matching order and ineffective pruning. It also introduced an efficient algorithm, \texttt{DAF}, for subgraph matching. In \cite{bhattarai2019ceci}, a Compact Embedding Cluster Index-based framework, \texttt{CECI}, has been proposed for subgraph listing. In \cite{sun2020memory}, a comparative analysis of various subgraph isomorphism algorithms including \texttt{QuickSI}, \texttt{GraphQL}, \texttt{CFL-Match}, \texttt{CECI}, \texttt{DP-iso}, \texttt{RI}, and \texttt{VF2++} on four different aspects (filtering techniques, matching orders, enumerating partial solutions, and other optimization techniques) is presented. The authors in \cite{sun2020subgraph} introduced a  novel approach to subgraph matching that considers the relationships between query vertices and the number of candidates, addressing limitations of path-based ordering and tree-structured indexing. Experimental results demonstrate significant performance improvements compared to existing methods, outperforming them by orders of magnitude on real-world and synthetic datasets.The authors in \cite{ning2020dominance} addressed subgraph matching in large RDF graphs by introducing a dominance-partitioned strategy that divides graphs into multiple clusters based on a dominance-connected pattern graph and employs a dominance-driven spectrum clustering approach, leading to improved time-efficiency for complex queries and enhanced scalability on various data scales and machines. The authors in \cite{shixuan2020RapidMatch} presented a subgraph query processing framework, \texttt{RapidMatch}, which is an integration of graph exploration and join based methods. Authors in \cite{mhedhbi2019optimizing} studied the problem of \textit{worst-case optimal join plan} for optimizing subgraph queries. The critical issue in the \textit{worst-case optimal join plan} is to get a matching order of query graphs. Recent advancements on graph pattern matching on massive graphs are presented in \cite{bouhenni2021survey}. Similarly, authors in \cite{mihelivc2021experimental} investigated many different filtering techniques used by the subgraph isomorphism solvers to speed up the searching process.
In the paper \cite{wang2022rdf} authors introduced an RDF subgraph query algorithm based on star decomposition to efficiently match subgraphs in large-scale RDF data, reducing communication costs, intermediate result sets, and improving query efficiency, as demonstrated through comparative experiments.
Authors in \cite{sun2022subgraph} focused on subgraph matching in knowledge graphs and introduced FGqT-Match, a novel algorithm that utilizes a subgraph index (FGqT) and a multi-label weight matrix to efficiently find all subgraph isomorphic mappings, outperforming existing algorithms in empirical studies on real and synthetic graphs.
In the  paper \cite{moayed2023efficient} authors presented  Eigen Decomposition Pruning (EDP), a novel approach for subgraph matching that leverages local spectral features, specifically Local Laplacian Matrices (LLMs), to reduce the search space, resulting in improved computational efficiency and reduced false positives. The proposed method is validated through theoretical analysis and experiments on real and synthetic datasets, demonstrating its effectiveness compared to existing methods in the field.

Constraint programming algorithms keep a list of candidate vertices that can map to each non-matched vertex of the query graph and then propagate edge and other constraints iteratively to reduce these lists. The main disadvantage of this approach is the amount of memory required to complete the matching. The \texttt{SND} (Scoring based Neighborhood Dominance) is a generalized filtering technique that may be used in subgraph matching algorithms to improve their performance \cite{audemard2014scoring}. It uses two parameters -- \textit{score} and \textit{neighborhood} functions. The \texttt{LAD} is a particular case of \texttt{SND} where filtering operation is based on local \texttt{alldifferent} constraints \cite{solnon2010alldifferent}. In this work, the authors modeled the subgraph matching problem into a constraint satisfaction framework and reported that their filtering technique improves the performance of the subgraph matching algorithms. The \texttt{Glasgow} subgraph matching algorithm supports both induced and non-induced variants of subgraph isomorphism problem wherein \textit{solution biased search} was introduced as a choice of backtracking search \cite{archibald2019sequential}. \texttt{Glasgow} uses the degree of a vertex to determine the proportion of effort for solving a subgraph isomorphism problem. \texttt{Glasgow} is sequential, but it can easily be converted into a parallel approach to speedup the matching process of the subgraph isomorphism problem. In \cite{mccreesh2018subgraph}, the authors proposed a graph generating method that generates ``really hard" instances of pattern and data graphs for induced and non-induced subgraph matching algorithms. The pattern and data graphs have a fixed number of vertices, and the pattern graphs of these instances have tens of edges, while the data graph contains a couple of hundred edges. They presented the results of \texttt{VF2}, \texttt{LAD}, and \texttt{Glasgow} on these generated instances of the pattern and data graph pairs. Authors of \cite{solnon2019experimental} extended the work of \cite{mccreesh2018subgraph} and compared subgraph matching algorithms \texttt{RI}, \texttt{VF3}, \texttt{Glasgow}, and \texttt{LAD} on broader data-sets. A method devolped to select subgraph isomorphism algorithm to handle hard instances subgraph isomorphism problem in \cite{kotthoff2016portfolios}. The selection depend on each problem instance not for entire query/pattern set.

\section{Preliminaries and Lemmas} \label{sec:BACKGROUND}

This section presents a formal definition of some of the basic concepts related to subgraph isomorphism problem. It also presents a formal description of the research problem addressed in this paper and proof of some basic lemmas that make the basis of our proposed $\mathrm{SubISO}$ algorithm. Table \ref{tab:Terminologiesused} summarizes the list of notations and their brief descriptions used in this paper.

\begin{defn}{\texttt{Graph:}}
A graph $G$ is a non-linear data structure, which can be defined as $G=<$\!$V, E, \Omega, l$\!$>$, where $V$ is the non-empty set of vertices, $E \subseteq V \times V$ is the set of edges, $\Omega$ is the set of vertex labels, and $l: V \rightarrow \Omega$ is a \textit{surjective} function which assigns a unique label from $\Omega $ to each vertex of $V$.
\end{defn}

\begin{figure}[!htb]
  \centering
  \subfigure[Query graph $G_{q}$]{\includegraphics[scale=0.38]{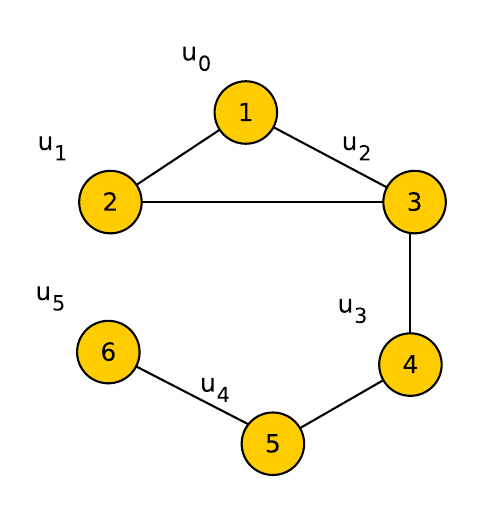}}\,
  \subfigure[Data graph $G_{d}$]{\includegraphics[scale=0.36]{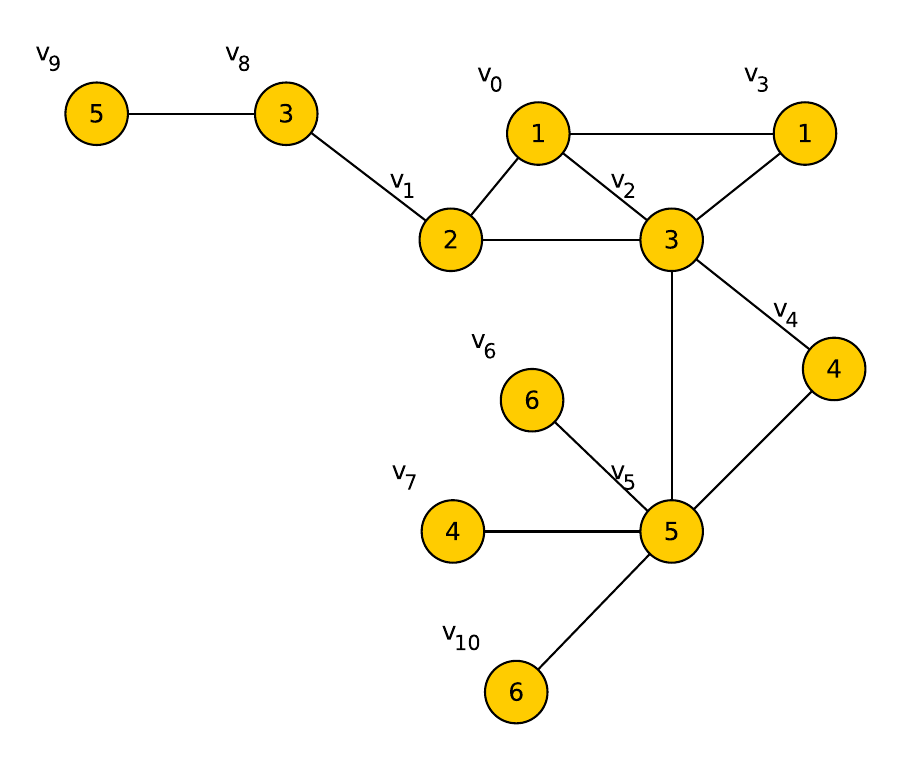}}
  \caption{An exemplar (a) query graph, and (b) data graph}
  \label{fig:ExemplerProblem}
\end{figure}

\begin{defn} {\texttt{Size of graph:}}
The size of a graph $G$ is defined as the number of vertices in its vertex set $V$, and it is denoted by $|V|$.
\end{defn}

\begin{defn}{\texttt{Subgraph:}}
A graph $G_{i}=~<$\!$V_{i}, E_{i}, \Omega_{i}, l_{i}$\!$>$ is a subgraph of another graph $G_{j} =~<$\!$V_{j}, E_{j}, \Omega_{j}, l_{j}$\!$>$ if $V_{i} \subseteq V_{j}$, $E_{i} \subseteq E_{j}$, $\Omega_{i}  \subseteq \Omega_{j}$, and $ l_{i}(u) =  l_{j}(u), \forall u \in V_{i}$ .
\end{defn}
\begin{defn}{\texttt{Degree:}}
The degree of a vertex $u \in V$ of $G$ is defined as the number of immediate neighbors of $u$ (i.e., vertices in $G$ that are directly connected to $u$). It is denoted by $\Delta(u)$.
\end{defn}

\begin{defn} {\texttt{Maximum neighborhood degree:}}
The maximum neighborhood degree of a vertex $u \in V$ of $G$ is denoted by $\tilde{\Delta}_{\mathcal{N}}(u)$, and it is defined as the highest degree of a vertex in the immediate neighborhood of $u$ in $G$.
\end{defn}

\begin{defn}{\texttt{Distance:}}
Given a pair of nodes $u, v \in V$ of $G$, the distance between $u$ and $v$ is denoted by $\delta(u,v)$ and defined as the length of the shortest path between $u$ and $v$ in $G$.
\end{defn}

\begin{defn} {\texttt{Eccentricity:}}
The eccentricity of a vertex $u \in V$ of a graph $G$ is denoted by $\epsilon (u)$, and it is defined as the maximum distance between $u$ and any other vertex of $G$. Mathematically, $\epsilon (u)=\underset{v \in V}{\max}\{\delta(u, v)\}$
\end{defn}

\begin{defn} {\texttt{$\epsilon$-neighborhood:}}
The $\epsilon$-neighborhood of a vertex $u \in V$ of a graph $G$ is denoted by $\mathcal{N}_\epsilon(u)$, and it is defined as the set of all those vertices of $G$ that are at maximum $\epsilon$-distance from $u$. Mathematically, $\mathcal{N}_\epsilon(u) = \{v \mid v \in V and~\delta(u, v) \leq \epsilon\}$
\end{defn}
\begin{table}[h]
  \begin{center}
  \caption{Notations and their descriptions}
	 	\label{tab:Terminologiesused}
	 	\scalebox{0.80}{
	 	\begin{tabular}{|c|l|}
    \hline
      Notation &  Description \\
       \hline
       $G$ & Graph  \\
        \hline
      $G_q$ & Query Graph  \\
        \hline
      $G_d$ & Data Graph  \\
        \hline
      $|V|$ & Size (no. of vertices) of a graph with vertex set $V$  \\
        \hline
        $\Delta(u)$ & Degree of vertex $u$\\
        \hline
        $l(u)$ & Label of vertex $u$\\
        \hline
        $\tilde{\Delta}_{\mathcal{N}}(u)$ & Maximum neighborhood degree \\
        \hline
        $\delta(u,v)$ & Distance between vertex $u$ and $v$\\
        \hline
        $\epsilon (u)$ & Eccentricity of vertex $u$\\
        \hline
        $\mathcal{N}_\epsilon(u)$ & $\epsilon$-neighborhood of vertex $u$\\
        \hline
        $\psi(u)$ & Set of candidate matches of vertex $u$ in $G_d$\\
        \hline
        $\hat{u}$ & $u$ as a pivot vertex\\
        \hline
        $\eta$ & Parameter to restrict maximum recursive calls\\
        \hline
        $\mathcal{R}$ & A candidate region in $G_d$\\
         \hline
        $\psi_\mathcal{R}(u)$ & Set of candidate matches of vertex $u$ in $\mathcal{R}$\\
        \hline
         $\mathpzc{M}$ & Embedding of $G_{q}$ in $G_{d}$\\
        \hline
        $\mathpzc{M_{R}}$ & Set of embeddings of $G_{q}$ in the candidate region  $\mathcal{R}$\\
        \hline
         $\mathpzc{M_{G_d}}$ & Set of embeddings of $G_{q}$ in $G_d$\\
         \hline
           \multirow{2}{*}{$\mathcal{C}(\mathcal{R})$}	& List of $<$\!$u,\psi(u)$\!$>$ in the candidate region $\mathcal{R}$, \\
         							&  $\forall u \in V(G_q)$ \\
          \hline
         \end{tabular} }
       
  \end{center}
\end{table}

\begin{defn} {\texttt{Subgraph isomorphism:}}\label{def:SubgraphIsomorphism}
	A graph $G_{p} =~<$\!$V_{p}, E_{p}, \Omega_{p}, l_{p}$\!$>$ is subgraph isomorphic to another graph $G_{q} =~<$\!$V_{q}, E_{q}, \Omega_{q}, l_{q}$\!$>$ if there exists an injective mapping $f: V(G_{p}) \rightarrow V(G_{q})$, which satisfies the following conditions:
	\begin{enumerate}
		\item $\forall u \in V_p, l_{p}(u) =  l_{q}(f(u))$
		\item $\forall (u,v) \in E_p$,  $\exists$ $(f(u),f(v)) \in E_q$.
	\end{enumerate}
\end{defn}

It may be noted that isomorphic image of a subgraph may contain extra edges, and it is generally referred as non-induced subgraph isomorphism.
  
\begin{defn} {\texttt{Isomorphic invariants:}}
Isomorphic invariants of a graph $G$ are the properties of $G$ that are preserved in isomorphic image of $G$.
\end{defn}

For example, number of vertices, number of edges, degree of a vertex, maximum degree of a vertex, and number of cycles, are some of the isomorphic invariants of a graph.

\begin{defn} {\texttt{Embedding:}}
Given a query graph $G_{q}$ and a data graph $G_{d}$, an embedding $\mathpzc{M}$ of $G_{q}$ in $G_{d}$ is the set of all ordered pairs $<$\!$u, v$\!$>$, where $u \in V(G_{q}) $, and $v \in V(G_d) $ is an subgraph isomorphic image of $u$.
\end{defn}

\begin{defn} {\texttt{Subgraph isomorphism problem:}}
For a given query graph $G_q$ and data graph $G_d$, the subgraph isomorphism problem is to enumerate all embeddings of $G_q$ in $G_d$.
\end{defn}

\texttt{Problem Statement:} For a given query graph $G_{q}$ and data graph $G_{d}$, we need to enumerate at most $k$ embeddings of $G_{q}$ in $G_{d}$.

In this study, we consider query graph as a labeled, connected, undirected, small in size graph, whereas data graph as a labeled, undirected, and large graph comprising million/billion vertices and edges. Figures \ref{fig:ExemplerProblem}(a) and \ref{fig:ExemplerProblem}(b) present an example of a query graph $G_{q}$ and data graph $G_d$, respectively, wherein the set of ordered pairs $\{<$\!$u_{0},v_{0}$\!$>,<$\!$u_{1},v_{1}$\!$>,<$\!$u_{2},v_{2}$\!$>,<$\!$u_{3},v_{4}$\!$>,<$\!$u_{4},v_{5}$\!$>,<$\!$u_{5},v_{6}$\!$>\}$ represents an embedding of $G_{q}$ in $G_d$. As stated earlier, the \textit{subgraph isomorphism problem} aims to enumerate all such embeddings of $G_{q}$ in $G_{d}$. However, in this study, we are interested to enumerate at most $k$ embeddings of $G_{q}$ in $G_{d}$ to deal with the \textit{output crisis} problem which has been discussed in \cite{qiao2017subgraph}.

\begin{lemma}\label{lemma:RegionSize}
Given the eccentricity $\epsilon$ of a vertex $u \in \mathit{V(G_q)}$ and a set of matching vertices of $u$ in $\mathit{G_d}$ $\psi(u)$, the maximum size of the candidate region corresponding to each vertex $v \in \psi(u)$ in $\mathit{G_d}$ is given by the cardinality of its $\epsilon$-neighborhood (i.e., $\abs{\mathcal{N}_\epsilon(v)}$).
\end{lemma}

\begin{proof}
Since $\mathit{G_q}$ is a connected graph and $\epsilon$ is the eccentricity of $u \in \mathit{V(G_q)}$, all other vertices of $\mathit{G_q}$ will be at most $\epsilon$ distance from $u$. Moreover, since distance between two vertices in a graph is an \textit{isomorphic invariant}, if $v \in \psi(u)$ is an isomorphic image of $u \in \mathit{V(G_q)}$ in $\mathit{G_d}$, then the isomorphic images of the other vertices of $\mathit{G_q}$ will be at most $\epsilon$ distance from $v$ in $\mathit{G_d}$. This implies that the isomorphic images of all vertices of $\mathit{G_q}$ must be a member of $\mathcal{N}_\epsilon(v)$. Hence, the maximum size of the candidate region corresponding to $v$ is given by $\abs{\mathcal{N}_\epsilon(v)}$.
\end{proof} 

It can be concluded from this lemma that the number of vertices in the candidate region of a vertex $u \in \mathit{G_q}$ is directly proportional to its eccentricity in $\mathit{G_q}$.

\begin{lemma}\label{lemma:computationcomplexity}
Maximum number of vertices that can be accessed in a data graph $\mathit{G_d}$ to find all subgraph isomorphisms of a query graph $\mathit{G_q}$ is given by $\mathcal{O}(\abs{\psi(u)} \times \epsilon)$, where $\psi(u)$ is the set of matching vertices of $u \in \mathit{G_q}$ in $\mathit{G_d}$, and $\epsilon$ is the eccentricity of $u$ in $\mathit{G_q}$. 
\end{lemma}

\begin{proof}
Given vertex $u$ and $ \psi(u) $, for each $v \in \psi(u) $ the maximum size of the respective candidate region in $\mathit{G_d}$ is $\abs{\mathcal{N}_\epsilon(v)}$. [Lemma \ref{lemma:RegionSize}]\\
Since $\forall v \in \psi(u)$, we need to explore one candidate region $\mathcal{N}_\epsilon(v)$, the maximum number of vertices to be explored to find all isomorphic images of $\mathit{G_q}$ in $\mathit{G_d}$ will be at most $\sum\nolimits_{v \in \psi(u)} \abs{\mathcal{N}_\epsilon(v)}$.\\
Since the size of a candidate region is directly proportional to $\epsilon$, we may write:\\ $\sum\nolimits_{v \in \psi(u)} \abs{\mathcal{N}_\epsilon(v)} \leq \sum\nolimits_{v \in \psi(u)} c_v \times \epsilon $, where $c_v$ is the proportionality constant depending on the size of $\mathcal{N}_\epsilon(v)$.\\ 
If $c_{max} = \max_{v \in \psi(u)} \{c_v$\}, then\\
$\sum\nolimits_{v \in \psi(u)} c_v \times \epsilon \leq \epsilon \times \sum\nolimits_{v \in \psi(u)} c_{max} = \epsilon \times c_{max} \times \abs{\psi(u)} $\\
$\Rightarrow \sum\nolimits_{v \in \psi(u)} \abs{\mathcal{N}_\epsilon(v)} \leq c_{max} \times \abs{\psi(u)} \times \epsilon $\\
Hence, $\sum\nolimits_{v \in \psi(u)} \abs{\mathcal{N}_\epsilon(v)} = \mathcal{O}(\abs{\psi(u)} \times \epsilon)$
\end{proof}

It may be noted that pivot vertex or starting vertex ($\hat{u}$) selection is one of the key steps in \textit{subgraph isomorphism problem} to enumerate all embeddings of the query graph into data graph efficiently. Therefore, based on lemmas \ref{lemma:RegionSize} and \ref{lemma:computationcomplexity}, we have designed the following objective function, which aims to minimize the total number of vertices that need to be accessed for enumerating all embeddings.\\
\begin{equation} \label{obj_fn}
\hat{u}=\underset{u \in \mathit{V(G_q)}}{\arg\min}\{\abs{\psi(u)} \times \epsilon(u)\}
\end{equation}
For a given query graph $ \mathit{G_q} $, this objective function is used to find the optimal pivot vertex of $ \mathit{G_q} $.

\section{Proposed Method} \label{sec:proposedmethod}

In this section, we present the procedural details of our proposed method, $\mathrm{Sub{ISO}}$, to enumerate all subgraphs of the data graph $G_d$ that are subgraph isomorphic to a given query graph $G_q$. Algorithm \ref{Algo:SubIso} presents the $\mathrm{SubISO}$ method formally which uses three major functions, namely $OptimalPivotVertexSelection()$, $RegionExploration()$, and $SubgraphSearch()$. The $OptimalPivotVert$ $exSelection()$ function is used to select an optimal pivot vertex ($\hat{u}$) of $G_q$, and a set of all candidate matches of ($\hat{u}$) in $G_d$ ($\psi(\hat{u})$) using the objective function defined in the previous section. The $RegionExploration()$ function finds a candidate region for each element of $\psi(\hat{u})$. Finally, the $SubgraphSearch()$ function is used to enumerate all embeddings of $G_q$ in the candidate regions of $G_q$.  Further details about these functions are presented in the following sub-sections.
 

\begin{algorithm}
\caption{$\mathrm{Sub{ISO}}(\mathit{G_q}, \mathit{G_d})$}\label{Algo:SubIso}
\begin{algorithmic}[1]
\Require $\mathit{G_q}$: query graph, $\mathit{G_d}$: data graph
\Ensure A set $\mathpzc{M_G}$ of all embeddings of $\mathit{G_q}$ in $\mathit{G_d}$ 

\State $<$\!$\hat{u},\psi(\hat{u})$\!$>\,\leftarrow \,$\textit{OptimalPivotVertexSelection}$(G_q,G_d)$
\State $\mathpzc{M_{G_d}} \leftarrow \phi$ \;
\State $\epsilon \leftarrow Eccentricity(\hat{u})$\;
\For{$v \in \psi(\hat{u}) $}
      \State $\mathcal{C}(\mathcal{R}) \leftarrow RegionExploration(\mathit{G_q},\mathit{G_d},\hat{u},\epsilon, v)$\;
      \If{$\mathcal{C}(\mathcal{R}) = \phi$}
	  \State 	\textbf{continue} \;
       \EndIf
      \State $\mathpzc{M} \leftarrow \phi$\;
	  \State $\mathpzc{M} \leftarrow \mathpzc{M} \cup \{<$\!$\hat{u},v$\!$>\}$\;
 	  \State $\mathpzc{M_{R}}\leftarrow SubgraphSearch(\mathit{G_q},\mathit{G_d},\mathcal{C}(\mathcal{R}),\mathpzc{M})$\;
	  \State $\mathpzc{M_{G_d}} \leftarrow \mathpzc{M_{G_d}}\cup \mathpzc{M_{R}} $\;
	\EndFor
\State \Return $\mathpzc{M_{G_d}}$\;	
\end{algorithmic}
\end{algorithm}
\subsection{Optimal Pivot Vertex Selection} \label{subsec:PivotVertexSelection}
As discussed earlier, the optimal pivot vertex of the query graph $G_q$ is a vertex in $G_q$ which minimizes the objective function defined in  equation \eqref{obj_fn}. It is used to explore candidate regions in $G_d$. It is also used as the starting vertex to identify subgraph isomorphisms of $G_q$ in the candidate regions of the data graph $G_d$. Algorithm \ref{Algo:PivotVertexSelection} presents the procedural details of the $OptimalPivotVertexSelection()$ function formally.

\begin{algorithm}
\caption{OptimalPivotVertexSelection$(\mathit{G_q}, \mathit{G_d})$}\label{Algo:PivotVertexSelection}
\begin{algorithmic}[1]
\Require $\mathit{G_q}$: query graph, $\mathit{G_d}$: data graph
\Ensure The pivot vertex $\hat{u}$ and its candidate matches $\psi(\hat{u})$

\For{ $u\in V(\mathit{G_q})$}
	\State $\widetilde{\psi}(u) \leftarrow \phi$\;
	\For{ $v \in V(\mathit{G_d})$}
     	\If{$l(u) = l(v)$ $\&$ $ \Delta(u) \leq \Delta(v)$ $\&$ $\tilde{\Delta}_{\mathcal{N}}(u) \leq \tilde{\Delta}_{\mathcal{N}}(v)$}
     		\State	$\widetilde{\psi}(u) \leftarrow  \widetilde{\psi}(u)\cup \{v\}$
		\EndIf    	
     \EndFor
\EndFor
\State $\mathcal{L} \leftarrow $Sort$(V(\mathit{G_q}))$\;
\rcomment{based on $\abs{\widetilde{\psi}(u)}$.}
\State    $\mathcal{L'} \leftarrow $ Select First Three member of $(\mathcal{L})$ \;
\For{ $u \in \mathcal{L'}$}
	\State $\psi(u) \leftarrow \phi$\;
	 \State $ l_{u}\leftarrow l(\mathcal{N}(u))$\;
	\For{ $v \in \widetilde{\psi}(u)$ }
     		\State $l_{v} \leftarrow l(\mathcal{N}(v))$\;
			\If{$l_{u} \subseteq l_{v}$}
		 			\State $\psi(u)\leftarrow  \psi(u)\cup \{v\}  $	
			\EndIf    		
	\EndFor
\EndFor
	
\State	$\hat{u} \leftarrow  \underset{u\in \mathcal{L'}}{\arg\min}\{\abs{\psi(u)}\times \epsilon(u)\}$ \;
	\rcomment{provided $\abs{\psi(u)}\times \epsilon(u) \neq 0 , \forall u \in \mathcal{L'} $}
	\State \Return $<$\!$\hat{u}$, $\psi(\hat{u})$\!$>$\;

\end{algorithmic}
\end{algorithm}

This algorithm uses three isomorphic  invariants -- \textit{label}, \textit{degree}, and \textit{maximum neighborhood degree} to find the set of matches, $\widetilde{\psi}(u)$, for every vertex $u$ of $G_q$. It uses the cardinality of $\widetilde{\psi}(u)$ to arrange the vertices of $G_q$ in ascending order and generates a set $\mathcal{L'}$ containing the first three vertices from the ordered list. Thereafter, for each element $u\in \mathcal{L'}$, it generates a list $l_u$ containing the labels of its immediate neighbors in $G_q$ and a list $l_v$ containing the labels of the immediate neighbors of the vertex $v \in \widetilde{\psi}(u)$ in $G_d$. If $l_u$ is contained in $l_v$, then vertex $v$ added in the set of candidate match $\psi(u)$ of the vertex $u$. This process is repeated for each $v \in \widetilde{\psi}(u)$. Finally, it chooses a vertex $u\in \mathcal{L'}$ as a pivot vertex, $\hat{u}$, which minimizes the product $\abs{\psi(u)}\times \epsilon(u)$. It may be noted that the product $\abs{\psi(u)}\times \epsilon(u)$ should be non-zero; otherwise, the query graph $G_q$ doesn't have any isomorphic image in the data graph $G_d$. The algorithm \ref{Algo:PivotVertexSelection} also returns a set, $\psi(\hat{u})$, containing all matching vertices of $\hat{u}$ in $G_d$.   

For example, consider a query graph $G_q $ and data graph $G_d$ given in figure \ref{fig:ExemplerProblem}(a) and \ref{fig:ExemplerProblem}(b). After following the steps 1-20 of the algorithm \ref{Algo:PivotVertexSelection}, $ \mathcal{L'}$ has three vertices $u_1$, $u_2$, and $u_3$ of $G_q$ with $ \psi(u_1)=\{v_1\}, \psi(u_2)=\{v_2\},$ and $ \psi(u_3)=\{v_4\}$. Similarly, eccentricity of the vertices $u_1$, $u_2$, and $u_3$ is 4, 3, and 2, respectively. Since $u_3$ minimize the product $\abs{\psi(u)}\times \epsilon(u)$, it can be selected as pivot vertex $\hat{u}$, and $\psi(\hat{u_3})=\{v_4\}$.

However, if we consider the rank function used in \texttt{$\mathrm{Turbo_{ISO}}$} \cite{han2013turbo} for pivot vertex selection, then $\hat{u} $ will be $u_1$, which is obviously not a good choice because the eccentricity of $u_1$ is 4. In this case, in order to match the farthest vertex of the query graph $G_q$, we need to explore the whole data graph as a candidate region. This will increase the number of recursive calls of the \texttt{SubgraphSearch()} function used in $\mathrm{Turbo_{ISO}}$. The \texttt{SubgraphSearch()} is a depth-first-search, recursive procedure of the Ullmann's method \cite{ullmann1976algorithm}. 

Similarly, if we choose pivot vertex using the rank function used in \texttt{CFL-Match} \cite{bi2016efficient}, then $\hat{u} $ will be $u_2$, which is also not a good choice because the eccentricity of $u_2$ is 3, resulting in increased matching process cost. Hence, instead of choosing $u_1$ or $u_2$ as pivot vertex, if we choose $u_3$ as pivot vertex (using algorithm \ref{Algo:PivotVertexSelection}), then it reduces the size of the candidate region from the whole data graph to the one shown in figure \ref{fig:CandidateRegion} (i.e., $(V(G_d) \smallsetminus \{v_8,v_9\})$), resulting in an efficient execution of the subgraph matching function.

\subsection{Region Exploration} \label{subsec:RegionExploration}
The region exploration process aims to find a candidate region around each matching vertex of the pivot vertex in the data graph $G_d$ using lemma \ref{lemma:RegionSize}. For a vertex $u \in G_d$, the candidate region, $\mathcal{R}$, is defined as the set of all those vertices of $G_d$ that are at most $\epsilon$-distance from $u$.   

\begin{algorithm}
\caption{RegionExploration$(\mathit{G_q},\mathit{G_d}, \epsilon, v)$} \label{Algo:RegionExploration}
\begin{algorithmic}[1]
\Require $\mathit{G_q}$: query graph, $\mathit{G_d}$: data graph, $\epsilon$: eccentricity of $\hat{u}$, $v$: a vertex of $\psi(\hat{u})$
\Ensure List $\mathcal{C}(\mathcal{R})$ of candidate matches of $\forall u \in G_q$ in the candidate region $\mathcal{R}$.

    \State $\mathcal{R}\leftarrow \mathcal{N}_\epsilon(v)$ \;
    \If{$\abs{\mathcal{R}} < ~\abs{V(\mathit{G_q})}$}
	  	\Return $\phi$  \;
    \EndIf    		
    \State $\mathcal{C}(\mathcal{R}) \leftarrow \phi$\;
    \rcomment{$\mathcal{C}(\mathcal{R})$ is a list of $<$\!$u,\psi_\mathcal{R}(u)$\!$>$}
    \State $\psi_\mathcal{R}(\hat{u}) \leftarrow \{v\}$\;
    \State $\mathcal{C}(\mathcal{R})\leftarrow \mathcal{C}(\mathcal{R}) \cup <$\!$\hat{u},\psi_\mathcal{R}(\hat{u})$\!$>$\;	
    \For{$u \in V(\mathit{G_q}) \smallsetminus {\hat{u}}$}
    \State 	$\psi_\mathcal{R}(u) \leftarrow \phi$\;
    \State 	$l_u \leftarrow l(\mathcal{N}(u))$\;
		\For{$ v \in \mathcal{R}$}
			\If{$l(u) = l(v)$ $\&$ $ \Delta(u) \leq \Delta(v)$ $\&$ $\tilde{\Delta}_{\mathcal{N}}(u) \leq \tilde{\Delta}_{\mathcal{N}}(v)$}
			 	\State $l_v \leftarrow l(\mathcal{N}(v))$\;
					\If{$l_u \subseteq l_v$}
						\State $\psi_\mathcal{R}(u)\leftarrow \psi_\mathcal{R}(u) \cup \{v\}$
					\EndIf
			\EndIf		
		\EndFor
		\If{$\psi_\mathcal{R}(u) = \phi $}
			\Return $ \phi $\;
		 \Else
    \State	$\mathcal{C}(\mathcal{R})\leftarrow \mathcal{C}(\mathcal{R}) \cup <$\!$u,\psi_\mathcal{R}(u)$\!$>$\;	
    	\EndIf
	\EndFor
	\State Sort $\mathcal{C}(\mathcal{R})$ in ascending order of $\frac{\abs{\psi_\mathcal{R}(u)}}{\Delta(u)}$\; \rcomment{Matching order.}
    \State \Return $\mathcal{C}(\mathcal{R})$\;
\end{algorithmic}
\end{algorithm}

It should be noted that if the size of a candidate region $\mathcal{R}$ is less than the size of the query graph, then it implies that there does not exist any embeddings of the query graph in $\mathcal{R}$, and it is simply ignored. On the other hand, if the size of $\mathcal{R}$ is greater than or equal to the size of the query graph, then the candidate region $\mathcal{R}$ may have an embedding of the query graph, and it needs to be preserved for further processing. To this end, for each query graph vertex $u \in G_q$, we find a set of its matches, $\psi_\mathcal{R}(u)$, in $\mathcal{R}$, and create a list $\mathcal{C(R)}$ containing all such $<$\!$u, \psi_\mathcal{R}(u)$\!$>$ ordered pairs. Since distance between any two vertices is an isomorphic invariant, only one match of $\hat{u}$ exists in $\mathcal{R}$. In order to find a match for each $u \in G_q$ in $\mathcal{R}$ except $\hat{u}$, we have used four isomorphic invariants -- \textit{label}, \textit{degree}, \textit{maximum neighborhood degree}, and \textit{label of immediate neighbors} of the query graph vertices. If should be noted that if $\psi_\mathcal{R}(u)$ for any node $u \in G_q$ in a region $\mathcal{R}$ is an empty set, then it is sufficient to decide that $\mathcal{R}$ will not have any embedding of $G_q$, and hence it is simply ignored from further processing.  

As discussed in \cite{bonnici2016variable}, the \textit{matching order} of vertices while finding subgraph isomorphism of a query graph plays an important role to accelerate the matching process. Accordingly, existing state-of-the-art methods like \texttt{QuickSI} and $\mathrm{Turbo_{ISO}}$ for finding subgraph isomorphisms have used different vertex matching orders. \texttt{QuickSI} \cite{shang2008taming} have used infrequent vertex labels and infrequent adjacent edge labels to determine matching order for the vertices of query graph. On the other hand, $\mathrm{Turbo_{ISO}}$ \cite{han2013turbo} have used a path-based ordering on query graph vertices while exploring the candidate region. However, both types of matching orders do not take into account the cost of the \texttt{SubgraphSearch()} function and the cost to check adjacency preservation between the vertices of the query graph and its corresponding images. Therefore, in line to the existing state-of-the-art methods, we have imposed a matching order on query graph vertices while exploring candidate regions. The cost of \texttt{SubgraphSearch()} function is directly proportional to $\abs{\psi_\mathcal{R}(u)}$, and the cost to check adjacency preservation is proportional to the degree of the vertex $u$ ($\Delta(u)$) in $G_q$. We have optimized both factors and ordered query graph vertices in ascending order based on the score of $\abs{\psi_\mathcal{R}(u)}/\Delta(u), \forall u \in G_q$. Our matching order gives higher preference to the vertex of query graph that has fewer candidate matches and a higher degree. 

\begin{figure}[!htb]
\begin{center}
\includegraphics[scale=0.38]{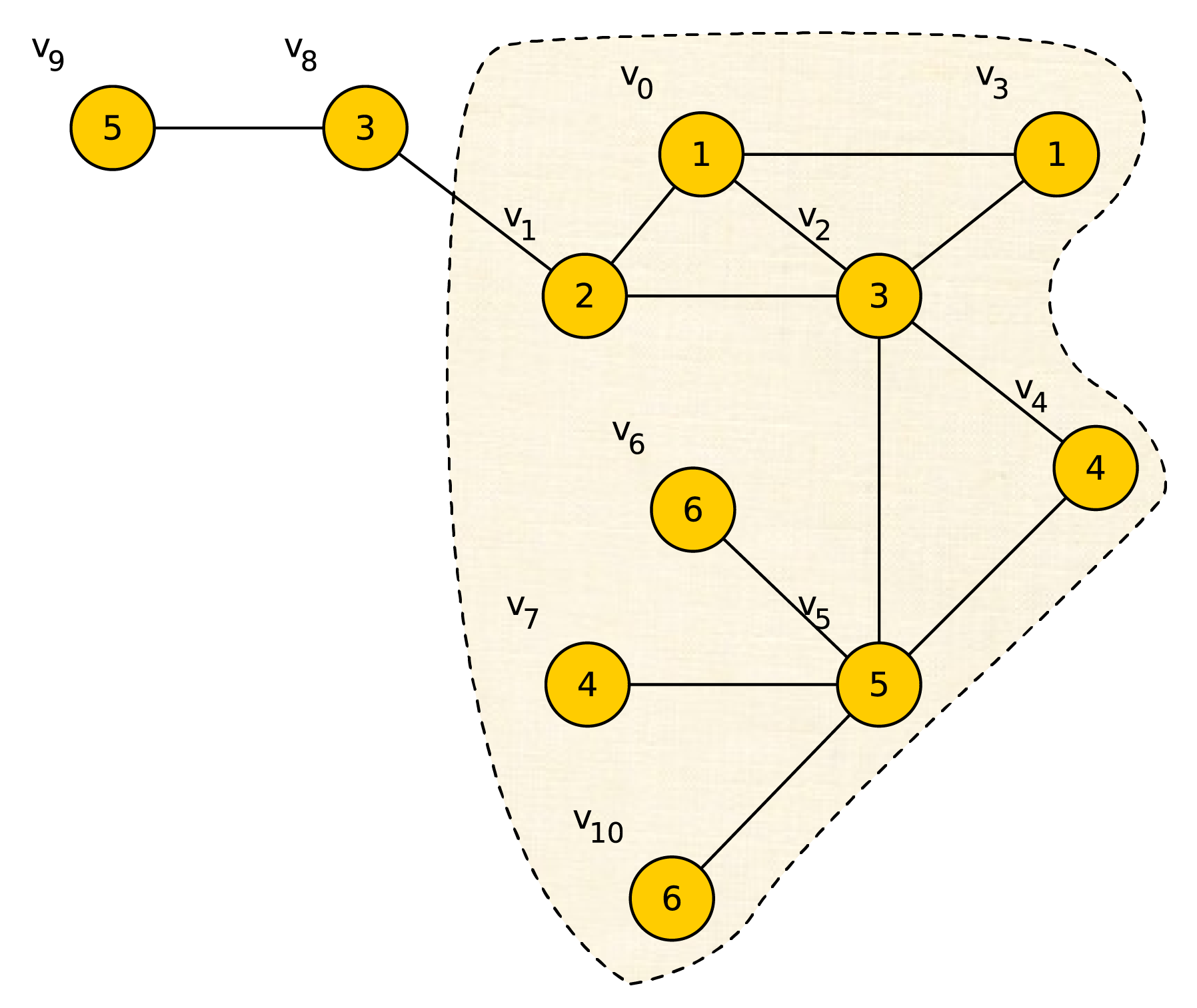} 
\caption{A candidate region ($\mathcal{R}$) in $G_d$ against the query graph $G_q$ shown in figure \ref{fig:ExemplerProblem}}
\label{fig:CandidateRegion}
\end{center}
\end{figure} 

Algorithm \ref{Algo:RegionExploration} presents the region exploration process discussed above formally.  Considering the same example of the data and query graphs, one identified candidate region $\mathcal{R}$ is shown in figure \ref{fig:CandidateRegion}. For this candidate region, the elements of the list $\mathpzc{C}(\mathcal{R})$ are $<$\!$u_0,\psi_\mathcal{R}(u_0)$\!$>$, $<$\!$u_1,\psi_\mathcal{R}(u_1)$\!$>$, $<$\!$u_2,\psi_\mathcal{R}(u_2)$\!$>$, $<$\!$u_3,\psi_\mathcal{R}(u_3)$\!$>$, $<$\!$u_4,\psi_\mathcal{R}(u_4)$\!$>$, $<$\!$u_5,\psi_\mathcal{R}(u_5)$\!$>$, and elements of each $\psi_\mathcal{R}(u_i)$ are as follows: $\psi_\mathcal{R}(u_{0})$ =  $ \{v_{0}\}$, $\psi_\mathcal{R}(u_{1})$ = $ \{v_{1}\}$, $\psi_\mathcal{R}(u_{2})$ = $ \{v_{2}\}$, $\psi_\mathcal{R}(u_{3})$ = $ \{v_{4}\}$, $\psi_\mathcal{R}(u_{4})$ = $ \{v_{5}\}$, and $\psi_\mathcal{R}(u_{5})$ = $ \{v_{6}, v_{10}\}.$

After sorting using the proposed matching order, the sorted list $\mathpzc{C}(\mathcal{R})$ is $\lbrace<$\!$u_2,\psi_\mathcal{R}(u_2)$\!$>$, $<$\!$u_0,\psi_\mathcal{R}(u_0)$\!$>$, $<$\!$u_1,\psi_\mathcal{R}(u_1)$\!$>$,$<$\!$u_3,\psi_\mathcal{R}(u_3)$\!$>$, $<$\!$u_4,\psi_\mathcal{R}(u_4)$\!$>$,$<$\!$u_5,\psi_\mathcal{R}(u_5)$\!$>\rbrace$. 

\subsection{Subgraph Enumeration} \label{subsec:SubgraphEnumeration}
After finding $\mathcal{C}(\mathcal{R})$ including the list of vertices and the set of candidate matches for every vertex of the query graph $G_q$ in $\mathcal{R}$ using algorithm \ref{Algo:RegionExploration}, the subgraph enumeration process aims to identify all embeddings of the query graph $G_q$ in the candidate region $\mathcal{R}$.

\begin{algorithm}
\caption{SubgraphSearch$(\mathit{G_q},\mathit{G_d},\mathcal{C}(\mathcal{R}),\mathpzc{M})$} \label{Algo:SubgraphSearch}
\begin{algorithmic}[1]
\Require $\mathit{G_q}$: query graph, $\mathit{G_d}$: data graph, $\mathcal{C}(\mathcal{R})$: list of vertices matches for query graph vertices in candidate region $\mathcal{R}$, $\mathpzc{M}$: list of embeddings
\Ensure List of all embeddings of the query graph $\mathit{G_q}$ in candidate region $\mathcal{R}$

\If{\textbar$\mathpzc{M}$\textbar = \textbar$ V(\mathit{G_q}) $\textbar}
           \State \textbf{Print}$(\mathpzc{M})$\;
        \EndIf		
        \State $u \leftarrow $ Choose an unmatched vertex of query graph $G_q$.\;
        \State $\psi_\mathcal{R}(u) \leftarrow \{$Consider all available candidate match of u from $\mathcal{C}(\mathcal{R})\}$\;     
        \For{Unmatched vertex $v \in \psi_\mathcal{R}(u)$}
        		\If{$IsJoinable(u, v, \mathpzc{M}, \mathit{G_q}, \mathit{G_d}) = True $}
            		\State $ \mathpzc{M} \leftarrow \mathpzc{M}\cup \{<$\!$u,v$\!$>\} $\;
            		\State SubgraphSearch$(\mathit{G_q},\mathit{G_d},\mathcal{C}(\mathcal{R}),\mathpzc{M})$\;
	            \State Delete$<$\!$u,v$\!$>$ from $\mathpzc{M}  $\;
            \EndIf			
  		\EndFor   
\end{algorithmic}  		                     
\end{algorithm}

In order to find all embeddings of the query graph $G_q$ in the candidate region $\mathcal{R}$, we have used \texttt{Subgraph} \texttt{Search()} function presented as algorithm \ref{Algo:SubgraphSearch} which is a depth-first-search procedure of the Ullmann's method \cite{ullmann1976algorithm}. Algorithm \ref{Algo:SubgraphSearch} uses recursion to get all possible embeddings of the query graph $G_q$ in candidate region $ \mathcal{R} $. To this end, it recursively adds query graph vertex $u$ and its possible match $v$ to an embedding $\mathpzc{M}$ if the node pair $<$\!$u, v$\!$>$ satisfies the adjacency condition described in algorithm \ref{Algo:IsJoinable}. Algorithm \ref{Algo:IsJoinable} checks adjacency condition with all previously added vertex pairs in $\mathpzc{M}$.

\begin{algorithm}
\caption{IsJoinable$(u, v, \mathpzc{M}, \mathit{G_q}, \mathit{G_d})$} \label{Algo:IsJoinable}
\begin{algorithmic}[1]
\Require Query vertex $u \in V(\mathit{G_q})$, candidate vertex $v \in \psi (u)$, $\mathpzc{M}$: embedding, $\mathit{G_q}$: query graph, $\mathit{G_d}$: data graph
\Ensure if $<$\!$u,v$\!$>$ joinable, return \texttt{True}, otherwise return \texttt{False}

  \For{$<$\!$u',v'$\!$> \in \mathpzc{M}$}
     \If{$(u,u') \in E(\mathit{G_q}) ~\&~ (v,v') \notin E(\mathit{G_d})$}
       \State \Return False\;
     \EndIf
   \EndFor   
\State \Return True\;
\end{algorithmic}  		                     
\end{algorithm}

\begin{table}[!htb]
  \begin{center}
	 \caption{Query graph embedding found in the data graph}
	 	\label{tab:identifiedEmbeddings}
	 	\scalebox{0.90}{
	 	\begin{tabular}{|c|c|}
    \hline
      S.No. & Embedding \\
       \hline
       1 & $<$\!$u_{2},v_{2}$\!$>$,$<$\!$u_{0},v_{0}$\!$>$,$<$\!$u_{1},v_{1}$\!$>$,$<$\!$u_{3},v_{4}$\!$>$,$<$\!$u_{4},v_{5}$\!$>$,$<$\!$u_{5},v_{6}$\!$>$\\
        \hline
       2 & $<$\!$u_{2},v_{2}$\!$>$,$<$\!$u_{0},v_{0}$\!$>$,$<$\!$u_{1},v_{1}$\!$>$,$<$\!$u_{3},v_{4}$\!$>$,$<$\!$u_{4},v_{5}$\!$>$,$<$\!$u_{5},v_{10}$\!$>$\\
        \hline
         \end{tabular} 
}
  \end{center}
\end{table}

\begin{figure}[!htb]
  \centering
  \subfigure[Embedding 1]{\includegraphics[scale=0.38]{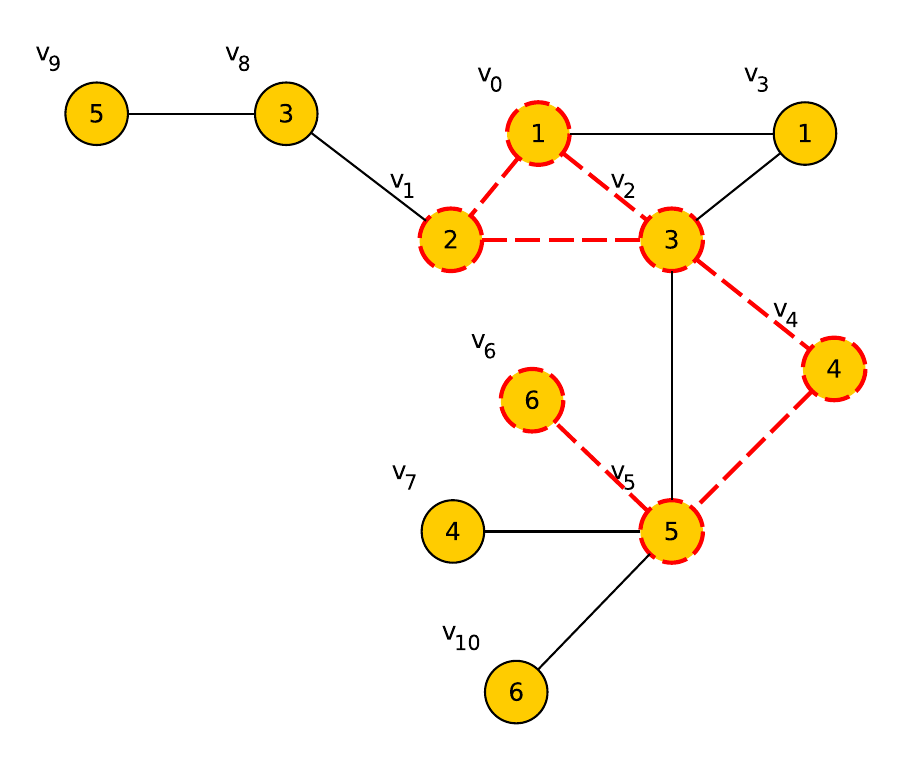}}\,
  \subfigure[Embedding 2]{\includegraphics[scale=0.38]{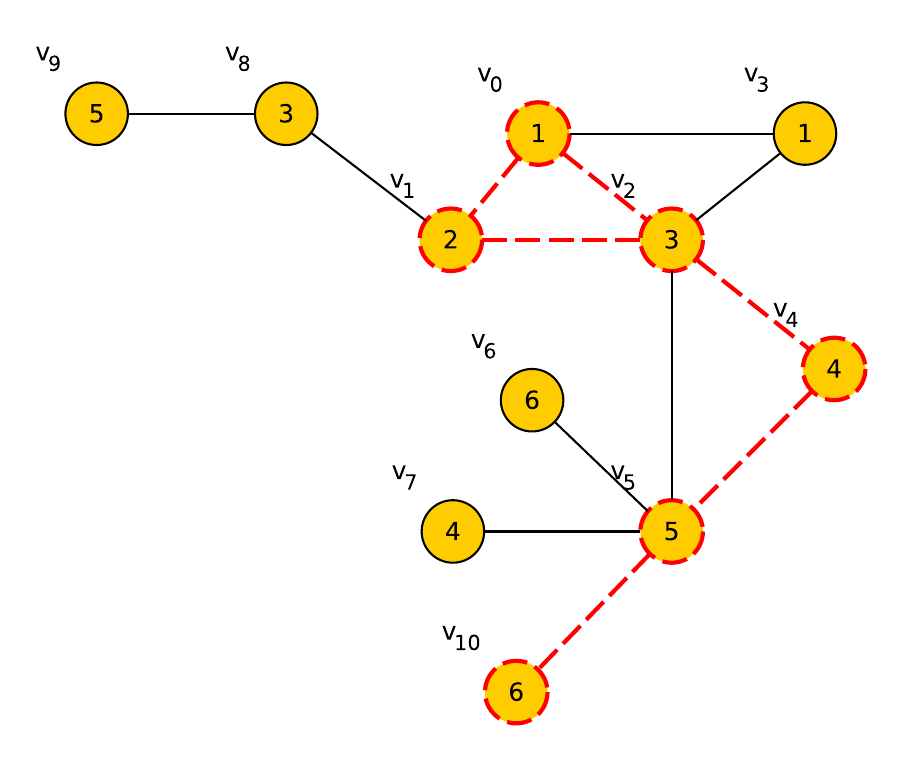}}
  \caption{Two embeddings of the query graph $G_q$ identified from the candidate region shown in figure \ref{fig:CandidateRegion}}
  \label{fig:IdentifiedEmbedding's}
\end{figure}

For example, two embeddings of the query graph $G_q$ identified from the candidate region $\mathcal{R}$ are listed in table \ref{tab:identifiedEmbeddings} and visualized in figures \ref{fig:IdentifiedEmbedding's}(a) and \ref{fig:IdentifiedEmbedding's}(b).

\section{Experimental Setup and Results} \label{sec:Experiments}
In this section, we present the experimental evaluation of the proposed $\mathrm{SubISO}$  method on different data and query graphs. All experiments are conducted on a PC having Intel Core i5-6600 processor and 4GB RAM. $\mathrm{SubISO}$ is implemented in C++ programming language using the \textit{igraph} library \cite{csardi2006igraph}, which has numerous functions to perform basic graph operations. In order to compile the $\mathrm{SubISO}$ program, we have used \texttt{g++} compiler on \texttt{Ubuntu 18.04.2 LTS} environment. Executable codes of $\mathrm{SubISO}$ and the datasets will be made public along with the publication of the manuscript.

\subsection{Data Graphs and Query Graphs} \label{subsec:DatasetandQuerySet}
In order to evaluate the performance of $\mathrm{SubISO}$, we have used three benchmark data graphs namely \texttt{Yeast}, \texttt{Human}, and \texttt{Hprd}. Table \ref{tab:DataGraphStatistics} presents the statistics of these data graphs. For each data graph, we have generated four query sets, each containing 100 simple, undirected, labeled, and connected graphs of a particular size (no. of vertices). We have taken query graph size as 10, 15, 20, and 25. Table \ref{tab:SelfQuerysetforYeastHumanHprd} presents the statistics of the query sets. In this table, $Yq_{n}$ represents the query set containing 100 query graphs of size $n$ (where $n \in \{10, 15, 20, 25\}$) for \texttt{Yeast} data graph. Similarly, other notations of this table can be interpreted. In order to create a query graph of a certain size, say $n$, we have randomly taken one node from the data graph and then traversed $n-1$ vertices of the data graph in breadth-first-search order. The resultant query graph is the labeled subgraph induced by $n$ traversed vertices of the data graph. We have repeated this process 100 times to create a query set of 100 such query graphs corresponding to a particular data graph. 

\begin{table}[!htb]
  \begin{center}
  \caption{Statistics of the data graphs}
  \scalebox{0.90}{
	 	\label{tab:DataGraphStatistics}
	 	\begin{tabular}{|l|c|c|c|c|}
    \hline
      Data graphs &  \#Vertices  &  \#Edges &  \#Distinct labels &  Avg degree \\
       \hline
       Yeast & 3112 & 12519 &  71 &  8.10  \\
        \hline
       Human & 4675 & 86282 &  44 &  36.82 \\
        \hline
       Hprd & 9460 & 34998 & 307 & 7.80 \\
        \hline
         \end{tabular} 
         }
  \end{center}
\end{table}

\begin{table}[!htb]
  \begin{center}
  \caption{Generated query sets for \texttt{Yeast}, \texttt{Human}, and \texttt{Hprd} data graphs}
	 	\label{tab:SelfQuerysetforYeastHumanHprd}
	 	\scalebox{0.90}{
	 	\begin{tabular}{|l|c|c|}
    \hline
      Data graph & Query set & Description \\
       \hline
       \multirow{3}{*}{Yeast}  & \multirow{3}{*}{$Yq_{n}$} & \multicolumn{1}{l|}{Query set $Yq_{n}$ contains 100}\\
     							&		&  \multicolumn{1}{l|}{query graphs of size $n$, }\\
        						&		&  \multicolumn{1}{l|}{where $n \in \{10,15,20,25\}$}\\
       
        \hline
       \multirow{3}{*}{Human}  & \multirow{3}{*}{$Hq_{n}$} & \multicolumn{1}{l|}{Query set $Hq_{n}$ contains 100}\\
     							&		&  \multicolumn{1}{l|}{query graphs of size $n$, }\\
        						&		&  \multicolumn{1}{l|}{where $n \in \{10,15,20,25\}$}\\

        \hline
        \multirow{3}{*}{Hprd}  & \multirow{3}{*}{$Pq_{n}$} & \multicolumn{1}{l|}{Query set $Pq_{n}$ contains 100}\\
     							&		&  \multicolumn{1}{l|}{query graphs of size $n$, }\\
        						&		&  \multicolumn{1}{l|}{where $n \in \{10,15,20,25\}$}\\
       
         \hline
        
         \end{tabular} 
         }
  \end{center}
\end{table}

We have also considered the query sets of one of the state-of-the-art methods, \texttt{CFL-Match} \cite{bi2016efficient}, that are generated for the \texttt{Human} data graph. There are total eight query sets for the \texttt{Human} data graph, and each query set contains 100 query graphs. Out of eight query sets, four have sparse query graphs and the remaining four have dense query graphs. These query sets are summarized in table \ref{tab:CFLQuerysetforHuman}. In this table, $CFLq_{nd}$ represents a query set of 100 dense query graphs each of size $n$ (where $n \in \{10, 15, 20, 25\}$). 

\begin{table}[!htb]
  \begin{center}
  \caption{CFL-Match's query sets for Human data graph}
	 	\label{tab:CFLQuerysetforHuman}
	 	\scalebox{0.90}{
	 	\begin{tabular}{|l|c|c|}
    \hline
      Data graph & Query set & Description \\
       \hline
 \multirow{6}{*}{Human}   & \multirow{3}{*}{$CFLq_{nd}$} & \multicolumn{1}{l|}{Query set $CFLq_{nd}$ contains }\\
         						&		& \multicolumn{1}{l|}{100 dense query graphs}\\
        						&		&\multicolumn{1}{l|}{of size $n$, where $n \in \{10,15,20,25\}.$}\\
       \cline{2-3}
         & \multirow{3}{*}{$CFLq_{ns}$} & \multicolumn{1}{l|}{Query set $CFLq_{ns}$ contains}\\
         						&		& \multicolumn{1}{l|}{100 sparse query graphs}\\
        						&		&\multicolumn{1}{l|}{of size $n$, where $n \in \{10,15,20,25\}.$}\\
       
        \hline
         \end{tabular} 
        }
  \end{center}
\end{table}

\subsection{Performance Evaluation Measures} \label{subsec:PEM}
In order to evaluate the performance of $\mathrm{SubISO}$, we have considered two well-known measures, namely the \textit{execution time} and the \textit{number of embeddings} found by a given algorithm. For a query set \textit{execution time} is calculated as the sum of the execution times to find at most $k$ embeddings for each query graph in the query set. On the other hand, the \textit{number of embeddings} for a query set is the sum of the embedding counts for each query graph in the query set. 

\subsection{Evaluation Results} \label{subsec:EvaluationResults}
This section presents the evaluation results of $\mathrm{SubISO}$ using our generated query sets and CFL-Match's query sets in terms of execution time and number of embeddings. Tables \ref{tab:Performance of  SubISO over data graphs Yeast, Human, Hprd} and \ref{tab:CFL Performance of  SubISO over data graps Human} present the evaluation results using our query sets and CFL-Match's query sets, respectively. For each query graph, we have set the value of the number of recursive calls $\eta$ parameter as 1000. Moreover, in line to the existing state-of-the-art methods like $\mathrm{Turbo_{ISO}}$ \cite{han2013turbo} and \texttt{QuickSI} \cite{shang2008taming}, the maximum number of embeddings per query graph to be printed is set to 1000.

\begin{table}[!htb]
  \begin{center}
  \caption{Performance evaluation results of $\mathrm{SubISO}$ using our generated query sets}
	 	\label{tab:Performance of  SubISO over data graphs Yeast, Human, Hprd}
	 	\scalebox{0.90}{
	 	\begin{tabular}{|l|c|c|c|}
    \hline
      Data graph & Q. set & \#Embedding & Exec. time (ms)\\
       \hline
       \multirow{4}{*}{Yeast} & $Yq_{10}$ & 65972 & 497\\  
       						  & $Yq_{15}$ & 65402 & 4287\\
       						  & $Yq_{20}$ & 68102 & 4960\\
       						  & $Yq_{25}$ & 69449 & 2680\\
        \hline
        \multirow{4}{*}{Human} & $Hq_{10}$ & 91386 & 6118\\  
       						   & $Hq_{15}$ & 84465 & 32796\\
       						   & $Hq_{20}$ & 80633 & 17895\\
       						   & $Hq_{25}$ & 86080 & 20024\\
        \hline
        \multirow{4}{*}{Hprd}  & $Pq_{10}$ & 5555 & 443\\  
       						   & $Pq_{15}$ & 9223 & 647\\
       						   & $Pq_{20}$ & 11524 & 910\\
       						   & $Pq_{25}$ & 19255 & 1177\\
        \hline
         \end{tabular} 
         }
  \end{center}
\end{table}

\begin{table}[!htb]
  \begin{center}
  \caption{Performance evaluation results of $\mathrm{SubISO}$ using CFL-Match's query sets}
  \scalebox{0.90}{
	 	\label{tab:CFL Performance of  SubISO over data graps Human}
	 	\begin{tabular}{|l|c|c|c|c|}
    \hline
      Data graph & Q. set & \#Embedding & Exec. time (ms)\\
       \hline
        \multirow{8}{*}{Human} & $CFLq_{10d}$ & 52623	& 2113\\  
       						   & $CFLq_{15d}$ & 41024	& 8732\\
       						   & $CFLq_{20d}$ & 20841	& 5900\\
       						   & $CFLq_{25d}$ & 19281	& 5327\\
       						   \cline{2-4}
       						   & $CFLq_{10s}$ & 60902	& 1403\\  
       						   & $CFLq_{15s}$ & 50850	& 5085\\
       						   & $CFLq_{20s}$ &	28186	& 4915\\
       						   & $CFLq_{25s}$ & 24606	& 9349\\
     \hline
         \end{tabular} 
         }
  \end{center}
\end{table} 

\subsection{Comparative Analysis} \label{subsec:ComparativeAnalysis}
In order to establish the efficacy of $\mathrm{SubISO}$ in comparison to the existing state-of-the-art methods, we have considered three well-known subgraph isomorphism algorithms, namely $\mathrm{Turbo_{ISO}}$ \cite{han2013turbo}, \texttt{QuickSI} \cite{shang2008taming} and \texttt{RI} \cite{bonnici2013subgraph}. As reported in \cite{lee2012depth}, \texttt{QuickSI} performs better than many other state-of-the-art methods for graph isomorphism. Therefore, we have not considered those algorithms for the comparative analysis of $\mathrm{SubISO}$. We have used the executable codes of $\mathrm{Turbo_{ISO}}$ and  \texttt{QuickSI} that are provided by the authors of \cite{han2013turbo} and \cite{lee2012depth}, respectively. Executable codes of \texttt{RI} \cite{bonnici2013subgraph} are available at \url{https://github.com/InfOmics/RI}. In \cite{sun2020memory}, the authors have shown that the overall performance of \texttt{RI} is comparable with other state-of-the-art methods over the \texttt{Yeast}, \texttt{Human}, and \texttt{Hprd} data graphs. We have used the default parameter settings of $\mathrm{Turbo_{ISO}}$, \texttt{RI}, and \texttt{QuickSI} in our experiments. It may be noted that both $\mathrm{Turbo_{ISO}}$ and \texttt{QuickSI} are set to print 1000 embeddings, by default, for each query graph. Therefore, we have run \texttt{RI} under the same setting to print 1000 embeddings.
 
\begin{figure}[!htb]
  \centering
  \subfigure[Yeast]{\includegraphics[scale=0.23,keepaspectratio]{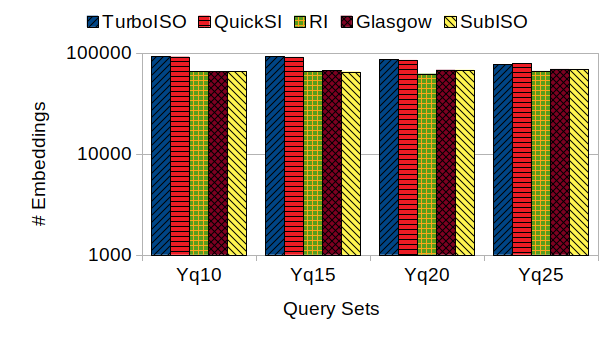}}\quad
  \subfigure[Human]{\includegraphics[scale=0.23,keepaspectratio]{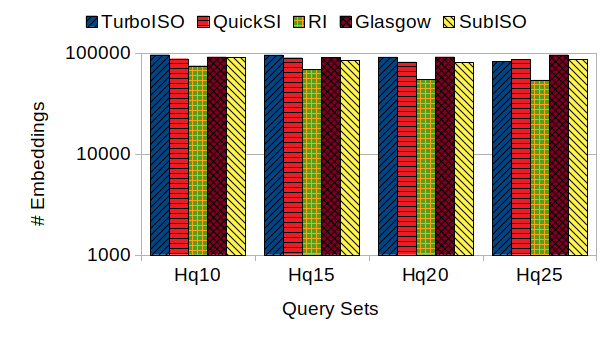}}\quad
   \subfigure[Hprd]{\includegraphics[scale=0.23,keepaspectratio]{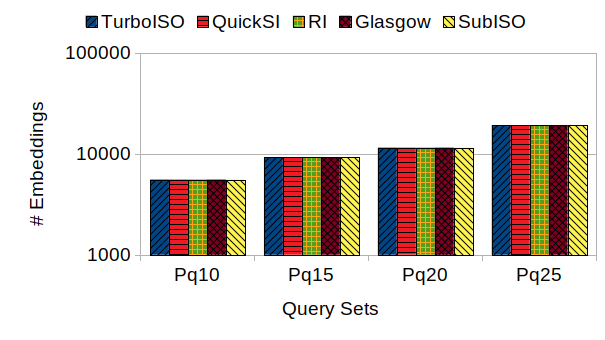}}
  \caption{Comparative analysis results of $\mathrm{Turbo_{ISO}}$, \texttt{QuickSI}, \texttt{RI}, \texttt{Glasgow} and $\mathrm{SubISO}$ in terms of number of embeddings for our generated query sets}
  \label{fig:ComparingNumberofEmbeddings}
\end{figure}

\begin{figure*}[!htb]
  \centering
  \subfigure[Yeast]{\includegraphics[scale=0.23,keepaspectratio]{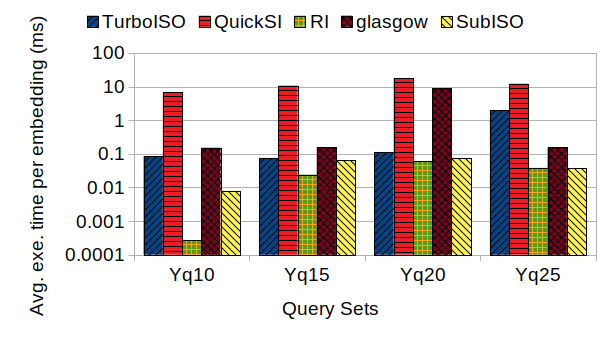}}\quad
  \subfigure[Human]{\includegraphics[scale=0.23,keepaspectratio]{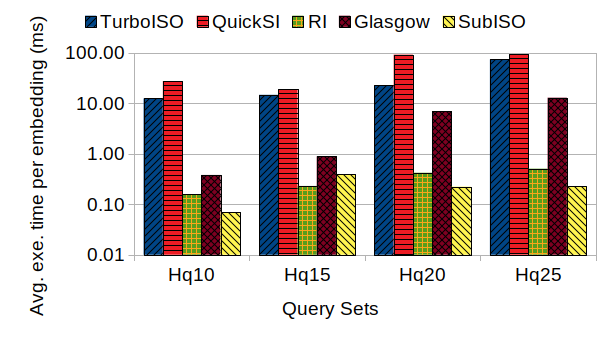}}\quad
   \subfigure[Hprd]{\includegraphics[scale=0.23,keepaspectratio]{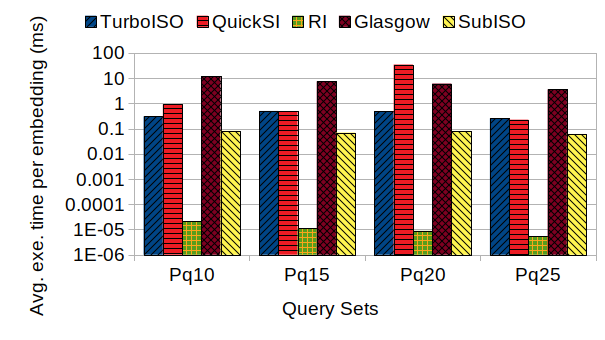}}
  \caption{Comparative analysis results of $\mathrm{Turbo_{ISO}}$, \texttt{QuickSI}, \texttt{RI}, \texttt{Glasgow} and $\mathrm{SubISO}$ in terms of average execution time per embedding for our generated query sets}
  \label{fig:ComparingExecutionTime}
\end{figure*}

Figures \ref{fig:ComparingNumberofEmbeddings} and \ref{fig:ComparingExecutionTime} present a comparison of  $\mathrm{SubISO}$ with $\mathrm{Turbo_{ISO}}$, \texttt{QuickSI} and \texttt{RI} in terms of the number of embeddings and average execution time per embedding for our generated query sets of the respective data graphs. Figure \ref{fig:ComparingNumberofEmbeddings} shows query sets at $x$-axis and number of embeddings identified by different algorithms on $y$-axis on a logarithmic scale. It can be observed from figure \ref{fig:ComparingNumberofEmbeddings}(a) that $\mathrm{SubISO}$ finds comparatively little-bit less number of embeddings on \texttt{Yeast} data graph. However, it is comparable to the state-of-the-art methods over the \texttt{Human} data graph, as shown in figure \ref{fig:ComparingNumberofEmbeddings}(b). Interestingly, all three algorithms find exactly same number of embeddings for all query sets over the \texttt{Hprd} data graph, as shown in figure  \ref{fig:ComparingNumberofEmbeddings}(c). Figure \ref{fig:ComparingExecutionTime} shows query sets at $x$-axis and average execution time per embedding of different algorithms on $y$-axis in milliseconds on a logarithmic scale. It can be observed from this figure that $\mathrm{SubISO}$ performs significantly faster in comparison to both $\mathrm{Turbo_{ISO}}$ and \texttt{QuickSI} over all data graphs.

\begin{figure}[!htb]
  \centering
  \subfigure{\includegraphics[scale=0.23,keepaspectratio]{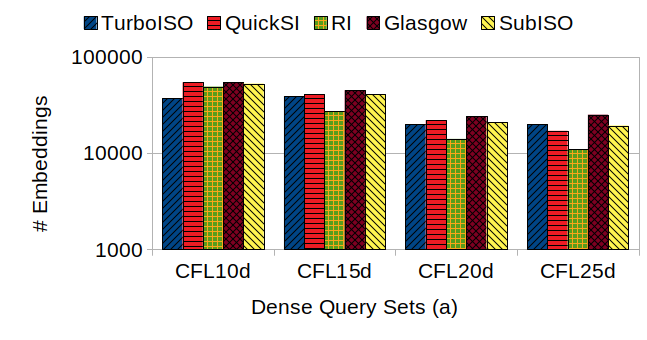}}\quad
  \subfigure{\includegraphics[scale=0.23,keepaspectratio]{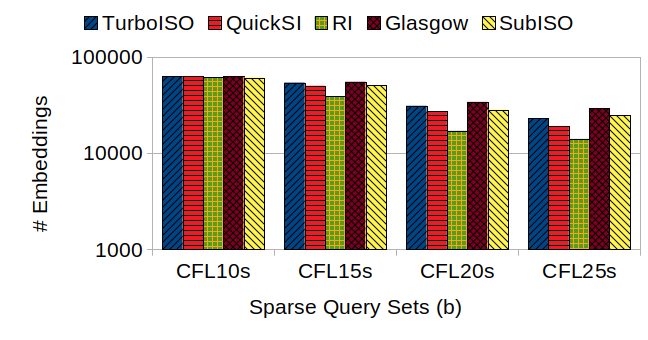}}
  \caption{Comparative analysis results of $\mathrm{Turbo_{ISO}}$, \texttt{QuickSI}, \texttt{RI}, \texttt{Glasgow} and $\mathrm{SubISO}$ in terms of number of embeddings for CFL-Match's query sets}
  \label{fig:CFLComparingNumberofEmbeddings}
\end{figure}
\begin{figure}[!htb]
  \centering
  \subfigure{\includegraphics[scale=0.23,keepaspectratio]{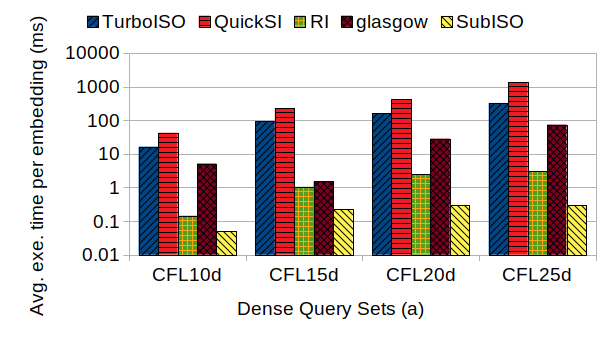}}\quad
  \subfigure{\includegraphics[scale=0.23,keepaspectratio]{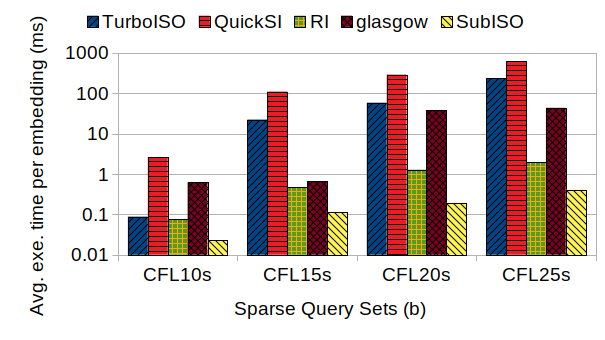}}
  \caption{Comparative analysis results of $\mathrm{Turbo_{ISO}}$, \texttt{QuickSI}, \texttt{RI}, \texttt{Glasgow} and $\mathrm{SubISO}$ in terms of average execution times per embedding for CFL-Match's query sets}
  \label{fig:CFLComparingExecutionTime}
\end{figure}


On the other hand, figures \ref{fig:CFLComparingNumberofEmbeddings} and \ref{fig:CFLComparingExecutionTime} present a comparison of  $\mathrm{SubISO}$ with $\mathrm{Turbo_{ISO}}$ and \texttt{QuickSI} in terms of the number of embeddings and execution time using CFL-Match's query sets of the \texttt{Human} data graph. Figure \ref{fig:CFLComparingNumberofEmbeddings} shows query sets at $x$-axis and number of embeddings identified by different algorithms on $y$-axis on a logarithmic scale. It can be observed from figure \ref{fig:CFLComparingNumberofEmbeddings} that $\mathrm{SubISO}$ is comparable to the state-of-the-art methods over both dense and sparse query sets. Figure \ref{fig:CFLComparingExecutionTime} shows query sets at $x$-axis and average execution time per embedding of different algorithms on $y$-axis in milliseconds on a logarithmic scale. It can be observed from this figure that, in this case too, $\mathrm{SubISO}$ performs significantly faster in comparison to both $\mathrm{Turbo_{ISO}}$ and \texttt{QuickSI} over both dense and sparse query sets. 

In figures \ref{fig:ComparingExecutionTime}(a) and \ref{fig:ComparingExecutionTime}(c) that are based on the \texttt{Yeast} and \texttt{hprd} data graphs, the execution time of \texttt{RI} is less than \texttt{SubISO}. On the other hand, in figures \ref{fig:ComparingExecutionTime}(b) and \ref{fig:CFLComparingExecutionTime} that are based on the \texttt{Human} data graph, the execution time of \texttt{SubISO} is less than \texttt{RI}. Since the structure of the \texttt{Human} data graph is complex in comparison to the \texttt{Yeast} and \texttt{hprd} data graphs, it can be deduced that \texttt{SubISO} outperforms \texttt{RI} over complex data graphs. 

Based on the aforementioned comparative analyses, it is found that $\mathrm{SubISO}$ is able to identify almost same number of embeddings in shorter time in comparison to the existing state-of-the-art methods. In other words, restricting the maximum number of recursive calls ($\eta$) makes $\mathrm{SubISO}$ computationally efficient to identify comparable number of embeddings.

\subsubsection{Dealing with Straggler Queries} \label{subsec:SQ}
For a given algorithm and data graph, it is still an open challenge to characterize straggler query, unless otherwise we execute the algorithm for a specified time-threshold. We have considered a query as \textit{straggler} query for an algorithm if the algorithm is not able to find any embedding even after executing it for 10 minutes ($6 \times 10^5$ milliseconds) as set in \cite{han2019efficient}, \cite{katsarou2018improving}. Through empirical analysis, we found that both $\mathrm{Turbo_{ISO}}$ and \texttt{QuickSI} did not find any embeddings for some query graphs even after running them for 5 hours. We have considered all such query graphs as straggler queries and performed a comparative analysis of $\mathrm{SubISO}$ to establish its efficacy in terms of dealing with straggler queries.

\begin{table}[!htb]
  \begin{center}
  \caption{Straggler queries for $\mathrm{Turbo_{ISO}}$ identified from our generated query sets }
  \scalebox{0.80}{
	 	\label{tab:StragglerqueriesofTurboISO}
	 	\begin{tabular}{|l|c|c|l|}
    \hline
      Data graph & Q. set & \#Straggler query & Q. graph id\\
       \hline
       \multirow{4}{*}{Yeast} & $Yq_{10}$ & 0 &  \\  
       						  & $Yq_{15}$ & 0 &  \\
       						  & $Yq_{20}$ & 0 &  \\
       						  & $Yq_{25}$ & 0 &  \\
        \hline
        \multirow{4}{*}{Human} & $Hq_{10}$ & 2 & 14, 22 \\  
       						   & $Hq_{15}$ & 2 & 39, 53 \\
       						   & $Hq_{20}$ & 3 & 19 , 31, 32\\
       						   & $Hq_{25}$ & 10 & 9, 23, 39, 47, 49, 55, 60, 64, 88, 93\\
        \hline
        \multirow{4}{*}{Hprd}  & $Pq_{10}$ & 0 &  \\
       						   & $Pq_{15}$ & 0 &  \\
       						   & $Pq_{20}$ & 0 &  \\
       						   & $Pq_{25}$ & 0 &  \\
        \hline
         \end{tabular} 
         }
  \end{center}
\end{table}

\begin{table}[!htb]
  \begin{center}
  \caption{Straggler queries for \texttt{QuickSI} identified from our generated query sets }
	 	\label{tab:StragglerqueriesofQuickSI}
	 	\scalebox{0.80}{
	 	\begin{tabular}{|l|c|c|l|}
    \hline
      Data graph & Q. set & \#Straggler query & Q. graph id\\
       \hline
       \multirow{4}{*}{Yeast} & $Yq_{10}$ & 1 & 85  \\  
       						  & $Yq_{15}$ & 1 & 91 \\
       						  & $Yq_{20}$ & 1 &  44 \\
       						  & $Yq_{25}$ & 1 & 12\\
        \hline
        \multirow{4}{*}{Human} & $Hq_{10}$ & 4 & 7, 13, 44, 57 \\  
       						   & $Hq_{15}$ & 2 & 24, 62 \\  
       						   & $Hq_{20}$ & 11 & 7, 11, 18, 29, 38, 45, 58, 77, 82, 87, 95\\
       						  & $Hq_{25}$ & 12 &  22, 31, 33, 41, 49, 50, 52, 55, 66, 79, 83, 96\\
        \hline
        \multirow{4}{*}{Hprd}  & $Pq_{10}$ & 0 &  \\
       						   & $Pq_{15}$ & 0 &  \\
       						   & $Pq_{20}$ & 0 &   \\
       						   & $Pq_{25}$ & 0 &  \\
        \hline

         \end{tabular} 
         }
  \end{center}
\end{table}

\begin{table}[!htb]
  \begin{center}
  \caption{Straggler queries for \texttt{RI} identified from our generated query sets}
	 	\label{tab:StragglerqueriesofRI}
	 	\scalebox{0.80}{
	 	\begin{tabular}{|l|c|c|l|}
    \hline
      Data graph & Q. set & \#Straggler query & Q. graph id\\
       \hline
       \multirow{4}{*}{Yeast} & $Yq_{10}$ & 0 &   \\  
       						  & $Yq_{15}$ & 1 & 62 \\
       						  & $Yq_{20}$ & 6 & 2, 13, 33, 48, 76, 91\\
       						  & $Yq_{25}$ & 4 & 37, 42, 73, 85\\
        \hline
        \multirow{10}{*}{Human}  & \multirow{2}{*}{$Hq_{10}$} & \multirow{2}{*}{17} & 7, 31, 32, 38, 44, 45, 48, 53, 68, 73, 77, 82, 86, 88, 92, \\  
       						   & &  & 93, 96\\ 
        \cline{2-4}
       						   & \multirow{2}{*}{$Hq_{15}$} & \multirow{2}{*}{22} & 2, 15, 21, 22, 30, 38, 39, 45, 47, 57, 59, 61, 62, 72, 73, \\ 
       						   & &  & 74, 75, 78, 85, 88, 92, 99\\ 
       	\cline{2-4}
       						   	& \multirow{3}{*}{$Hq_{20}$} & \multirow{3}{*}{37} & 1, 3, 4, 5, 7, 11, 14, 16, 20, 22, 24, 28, 30, 31, 32, 34,\\
       						   & &  & 35, 36, 38, 42, 45, 52, 54, 56, 67, 69, 71, 72, 73, 82,\\
       						   & &  & 83, 91, 94, 95, 97, 98, 99\\
       	\cline{2-4}
       						 		   & \multirow{3}{*}{$Hq_{25}$} & \multirow{3}{*}{44} & 2, 4, 5, 9, 16, 18, 23, 24, 25, 27, 28, 29, 30, 31, 33, 35,\\
       						   & &  & 36, 43, 44, 48, 49, 51, 55, 57, 58, 59, 60, 61, 66, 68,\\ 
       						   & &  & 71, 73, 77, 78, 79, 83, 86, 87, 90, 93, 95, 96, 97, 98\\ 
       \hline

        \multirow{4}{*}{Hprd}  & $Pq_{10}$ & 0 &   \\
       						   & $Pq_{15}$ & 0 &   \\
       						   & $Pq_{20}$ & 0 &   \\
       						   & $Pq_{25}$ & 0 &   \\
        \hline

         \end{tabular} 
         }
   
  \end{center}
\end{table}

\begin{table}[!htb]
  \begin{center}
  \caption{Straggler queries for \texttt{Glasgow} identified from our generated query sets}
	 	\label{tab:StragglerqueriesofGlasgow}
	 	\scalebox{0.80}{
	 	\begin{tabular}{|l|c|c|l|}
    \hline
      Data graph & Q. set & \#Straggler query & Q. graph id\\
       \hline
       \multirow{4}{*}{Yeast} & $Yq_{10}$ & 0 &   \\  
       						  & $Yq_{15}$ & 0 &  \\
       						  & $Yq_{20}$ & 1 & 35\\
       						  & $Yq_{25}$ & 0 & \\
        \hline
         \multirow{4}{*}{Human} & $Hq_{10}$ & 0 &   \\  
       						    & $Hq_{15}$ & 0 &  \\
       						    & $Hq_{20}$ & 1 & 18\\
       						    & $Hq_{25}$ & 2 & 66, 79\\
       \hline

        \multirow{4}{*}{Hprd}  & $Pq_{10}$ & 0 &   \\
       						   & $Pq_{15}$ & 0 &   \\
       						   & $Pq_{20}$ & 0 &   \\
       						   & $Pq_{25}$ & 0 &   \\
        \hline

         \end{tabular} 
         }
   
  \end{center}
\end{table}

\begin{table}[!htb]
  \begin{center}
  \caption{Straggler queries for $\mathrm{Turbo_{ISO}}$ identified from CFL-Match's query sets}
  \scalebox{0.80}{
	 	\label{tab:CFLStragglerqueriesofTurboISO}
	 	\begin{tabular}{|l|c|c|l|}
    \hline
      Data graph & Q. set & \#Straggler query & Q. graph id\\
       
        \hline
        \multirow{8}{*}{Human} & $CFLq_{10s}$ & 0 & \\  
       						   & $CFLq_{15s}$ & 2 &  75, 81\\
       						   & $CFLq_{20s}$ & 3 &  32, 67, 98\\
       						   & $CFLq_{25s}$ & 9 &  4, 5, 9, 14, 24, 27, 35, 41, 74\\
       						   & $CFLq_{10d}$ & 1 &  33\\  
       						   & $CFLq_{15d}$ & 6 &  9, 41, 42, 54, 78, 88\\
       						   & $CFLq_{20d}$ & 5 &  30, 49, 55, 65, 82 \\
       						   & $CFLq_{25d}$ & 10 &  0, 2, 7, 27, 30, 50, 58, 65, 73, 94\\
       						   
        \hline
         \end{tabular} 
         }
  \end{center}
\end{table}

\begin{table}[!htb]
  \begin{center}
  \caption{Straggler queries for \texttt{QuickSI} identified from \texttt{CFL-Match}'s query sets}
   \scalebox{0.80}{
	 	\label{tab:CFLStragglerqueriesofQuickSI}
	 	\begin{tabular}{|l|c|c|l|}
    \hline
      Data graph & Q. set & \#Straggler query & Q. graph id\\
       \hline
       \multirow{8}{*}{Human} & $CFLq_{10s}$ & 0 &  \\  
       						   & $CFLq_{15s}$ & 8	& 7, 18, 51, 72, 74, 75, 78, 94\\
       						   & $CFLq_{20s}$ & 12	& 15, 31, 32, 35, 51, 60, 65, 69, 71, 78, 85, 97\\
       						   & $CFLq_{25s}$ & 19	& 0, 4, 5, 7, 9, 14, 16, 17, 25, 27, 35, 41, 61, 68, 77, 78, 79, 88, 99\\
       						   & $CFLq_{10d}$ & 2	& 30, 76\\  
       						   & $CFLq_{15d}$ & 14	& 2, 9, 22, 25, 26, 38, 45, 46, 57, 66, 77, 82, 86, 89\\
       						   & $CFLq_{20d}$ & 15	& 13, 15, 23, 30, 32, 34, 38, 40, 56, 57, 62, 70, 83, 90, 95 \\
       						   & $CFLq_{25d}$ & 16	& 0, 2, 7, 12, 14, 22, 27, 29, 30, 55, 58, 67, 71, 73, 78, 94\\
        \hline
         \end{tabular} 
         }
  \end{center}
\end{table}

\begin{table}[!htb]
  \begin{center}
  \caption{Straggler queries for \texttt{RI} identified from \texttt{CFL-Match}'s query sets}
   \scalebox{0.80}{
	 	\label{tab:CFLStragglerqueriesofRI}
	 	\begin{tabular}{|l|c|c|l|}
    \hline
      Data graph & Q. set & \#Straggler query & Q. graph id\\
       \hline
       \multirow{20}{*}{Human} & $CFLq_{10s}$ & 5 &  4, 25, 44, 53, 62\\
       							\cline{2-4}  
       						   & \multirow{2}{*}{$CFLq_{15s}$} & \multirow{2}{*}{26}	& 0, 3, 5, 7, 14, 15, 19, 29, 31, 32, 36, 42, 43, 45, 51, 55,\\
       						    & &  & 63, 69, 70, 75, 77, 80, 81, 89, 95, 98\\
       						   \cline{2-4}
       						   & \multirow{3}{*}{$CFLq_{20s}$} & \multirow{3}{*}{35}	&  1, 6, 8, 11, 12, 15, 18, 20, 21, 28, 31, 32, 34, 37, 41, 42,\\
       						   & &  & 43, 45, 47, 48, 53, 60, 64, 68, 69, 71, 73, 75, 76, 85, 95,\\
       						   & &  & 96, 97, 98, 99\\
       						     \cline{2-4}
       						   & \multirow{3}{*}{$CFLq_{25s}$} & \multirow{3}{*}{44}	& 3, 4, 7, 9, 12, 13, 14, 15, 16, 19, 21, 24, 25, 27, 28, 29,\\
       						   & &  & 33, 34, 35, 38, 41, 54, 55, 58, 61, 62, 65, 67, 68, 69, 72,\\
       						   & &  & 74, 77, 78, 79, 80, 82, 83, 88, 89, 91, 94, 98, 99\\
       						     \cline{2-4}
       						   & $CFLq_{10d}$ & 9	& 8, 10, 13, 23, 30, 38, 60, 65, 80\\  
       						   \cline{2-4}
       						   & \multirow{3}{*}{$CFLq_{15d}$} & \multirow{3}{*}{42}	& 0, 1, 2, 5, 7, 8, 9, 11, 12, 14, 16, 17, 20, 21, 22, 24, 25,\\
       						    & &  & 34, 37, 38, 39, 40, 41, 45, 46, 48, 57, 59, 65, 66, 70, 74,\\
       						    & &  & 77, 78, 81, 82, 83, 86, 89, 92, 94, 98\\
       						   \cline{2-4}
       						   & \multirow{4}{*}{$CFLq_{20d}$} & \multirow{4}{*}{54}	& 1, 2, 3, 6, 8, 11, 15, 16, 17, 18, 20, 22, 23, 27, 30, 32, 33, \\
       						    & &  & 36, 39, 41, 42, 43, 44, 45, 46, 47, 48, 51, 54, 55, 57, 59,\\
       						    & &  & 62, 64, 65, 67, 70, 74, 75, 76, 78, 79, 81, 82, 83, 84,\\
       						    & &  & 89, 90, 92, 94, 95, 97, 99\\
       						
       						   \cline{2-4}
       						   & \multirow{4}{*}{$CFLq_{25d}$} & \multirow{4}{*}{54}	& 0, 2, 3, 4, 9, 10, 12, 14, 20, 22, 23, 26, 27, 28, 29,  30,\\
       						   	 & &  & 31, 34, 37, 38, 39, 40, 41, 42, 43, 44, 46, 47, 48, 49, 50,\\
       						   	 	& &  & 51, 52, 53, 55, 56, 57, 58, 59, 60, 63, 64, 65, 67, 71,\\
       						   	 	& &  & 72, 79, 81, 83, 90, 92, 93, 96, 99\\
        \hline
         \end{tabular} 
         }
  \end{center}
\end{table}

\begin{table}[!htb]
  \begin{center}
  \caption{Straggler queries for \texttt{Glasgow} identified from \texttt{CFL-Match}'s query sets}
   \scalebox{0.80}{
	 	\label{tab:CFLStragglerqueriesofGlasgow}
	 	\begin{tabular}{|l|c|c|l|}
    \hline
      Data graph & Q. set & \#Straggler query & Q. graph id\\
       \hline
       \multirow{8}{*}{Human}  & $CFLq_{10s}$ & 0 &  \\  
       						   & $CFLq_{15s}$ & 0	& \\
       						   & $CFLq_{20s}$ & 0	& \\
       						   & $CFLq_{25s}$ & 2	& 78, 89\\
       						   & $CFLq_{10d}$ & 0	& \\  
       						   & $CFLq_{15d}$ & 0	& \\
       						   & $CFLq_{20d}$ & 1	& 70 \\
       						   & $CFLq_{25d}$ & 3	& 10. 58, 73\\
        \hline
         \end{tabular} 
         }
  \end{center}
\end{table}

Tables \ref{tab:StragglerqueriesofTurboISO}, \ref{tab:StragglerqueriesofQuickSI}, and \ref{tab:StragglerqueriesofRI} present the details of the straggler queries for $\mathrm{Turbo_{ISO}}$, \texttt{QuickSI}, and \texttt{RI}, respectively that are identified from our query sets generated for the \texttt{Yeast}, \texttt{Human}, and \texttt{Hprd} data graphs. Similarly, Tables \ref{tab:CFLStragglerqueriesofTurboISO}, \ref{tab:CFLStragglerqueriesofQuickSI}, and \ref{tab:CFLStragglerqueriesofRI} present the details of the straggler queries for $\mathrm{Turbo_{ISO}}$, \texttt{QuickSI}, and \texttt{RI}, respectively that are identified from the \texttt{CFL-Match}'s query sets for the \texttt{Human} data graph. It can be observed from these tables that $\mathrm{Turbo_{ISO}}$ has less number of straggler queries in comparison to the \texttt{QuickSI} and \texttt{RI}. Moreover, each algorithm has a different, though overlapping, set of the straggler queries.

\begin{table}[!htb]
  \begin{center}
  \caption{Performance comparison of $\mathrm{SubISO}$ to deal with $\mathrm{Turbo_{ISO}}$'s straggler queries identified from our generated query sets}
   \scalebox{0.80}{
	 	\label{tab:Statistics of TurboISO Straggler queries1}
	 	\begin{tabular}{|l|c|c|c|c|c|}	 
    \hline
    \multirow{2}{*}{Query set/Data graph} & \multirow{2}{*}{Q. graph id} & \multicolumn{2}{c|}{TurboISO} & %
    \multicolumn{2}{c|}{SubISO}\\
\cline{3-6}
 & &  \#Embedding & Exec. time(ms) & \#Embedding & Exec. time(ms)\\
    	\hline       	 			    	
       	 \multirow{2}{*}{$Hq_{10}$/Human} & 14  & None & $6\times10^{5}$ & 1000 & 4 \\
       	 									& 22  & None & $6\times10^{5}$ & 1000 & 4 \\  
       	\hline
         \multirow{2}{*}{$Hq_{15}$/Human} & 39  & None & $6\times10^{5}$ & 1000 & 4 \\  
       						    			& 53  & None & $6\times10^{5}$ & 1000 & 4 \\	
       	 \hline					    	
       	 \multirow{2}{*}{$Hq_{20}$/Human} & 19  & None & $6\times10^{5}$ & 1000 & 140\\  	
       	 									& 32  & None & $6\times10^{5}$ & 1000 & 5 \\    
       	 										  	
       	 	\hline
         \multirow{2}{*}{$Hq_{25}$/Human} & 23  & None & $6\times10^{5}$ & 1000 & 11 \\    
       						    			& 60  & None & $6\times10^{5}$ & 1000 & 36\\								    							    			 				    								    			
       	\hline					    			
         \end{tabular} 
       }
  \end{center}
\end{table}

\begin{table}[!htb]
  \begin{center}
  \caption{Performance comparison of $\mathrm{SubISO}$ to deal with $\mathrm{Turbo_{ISO}}$'s straggler queries identified from \texttt{CFL-Match}'s query sets}
  \scalebox{0.80}{
	 	\label{tab:Statistics of TurboISO Straggler queries0}
	 	\begin{tabular}{|l|c|c|c|c|c|}	 
    \hline
    \multirow{2}{*}{Query set/Data graph} & \multirow{2}{*}{Q. graph id} & \multicolumn{2}{c|}{TurboISO} & %
    \multicolumn{2}{c|}{SubISO}\\
\cline{3-6}
 & &  \#Embedding & Exec. time(ms) & \#Embedding & Exec. time(ms)\\
    	\hline
    	\multirow{1}{*}{$CFLq_{15s}$/Human} & 81  & None & $6\times10^{5}$ & 954 & 79\\
       \hline
        \multirow{2}{*}{$CFLq_{20s}$/Human} & 67  & None & $6\times10^{5}$ & 1000 & 6\\
       						    			& 98  & None  & $6\times10^{5}$ & 1000  & 186\\  
        \hline
         \multirow{2}{*}{$CFLq_{25s}$/Human} & 4  & None & $6\times10^{5}$ & 1000	& 6006\\  
       						    			& 24  & None & $6\times10^{5}$ & 969 &	16 \\
       \hline
         \multirow{1}{*}{$CFLq_{10d}$/Human} & 33  & None & $6\times10^{5}$ & 1000 & 7\\  
       	\hline
         \multirow{2}{*}{$CFLq_{15d}$/Human} & 42  & None & $6\times10^{5}$ & 972	&  10\\    
       						    			& 54  & None & $6\times10^{5}$ & 1000	& 10\\					    			 
        \hline
         \multirow{2}{*}{$CFLq_{20d}$/Human} & 49  & None & $6\times10^{5}$ & 1000 & 19\\    
       						    			& 65  & None & $6\times10^{5}$  & 1000 & 12\\
        \hline
         \multirow{2}{*}{$CFLq_{25d}$/Human} & 2  & None & $6\times10^{5}$ & 1000	& 95\\    
       						    			& 94  & None & $6\times10^{5}$ & 1000	& 163\\
       	 \hline		
 \end{tabular} 
       }
  \end{center}
\end{table}       	 
       	 
Tables \ref{tab:Statistics of TurboISO Straggler queries1} and \ref{tab:Statistics of TurboISO Straggler queries0} present the performance comparison results of $\mathrm{SubISO}$ in terms of \texttt{number of embeddings} and \texttt{execution time} to deal with $\mathrm{Turbo_{ISO}}$'s straggler queries identified from our generated query sets (table \ref{tab:StragglerqueriesofTurboISO}) and CFL-Match's query sets (table \ref{tab:CFLStragglerqueriesofTurboISO}), respectively. We have randomly taken maximum two queries from each query set for this comparative evaluation. For each query, we have set upper time limit as $6\times10^{5}$ milliseconds (10 minutes), and recorded the number of embeddings found and the respective execution times when the algorithm terminates. If we did not find any output during the maximum time allowed, then we have noted ``none" as the \texttt{number of embeddings} and $6\times10^{5}$ milliseconds (10 minutes) as the \texttt{execution time}. In case of $\mathrm{SubISO}$, we have set the value of $\eta$ as 1000 and the maximum number of embeddings to be found as 1000, which is the default setting of $\mathrm{Turbo_{ISO}}$. It can be observed from these tables that $\mathrm{SubISO}$ finds embeddings for every $\mathrm{Turbo_{ISO}}$'s straggler queries in a very short execution time.

\begin{table}[!htb]
  \begin{center}
  \caption{Performance analysis of $\mathrm{SubISO}$ to deal with \texttt{QuickSI's} straggler queries identified from our generated query sets}
  \scalebox{0.80}{
	 	\label{tab:Statistics of QuickSI Straggler queries1}
	 		\begin{tabular}{|l|c|c|c|c|c|}	 
    \hline
    \multirow{2}{*}{Query set/Data graph} & \multirow{2}{*}{Q. Graph id} & \multicolumn{2}{c|}{QuickSI} & %
    \multicolumn{2}{c|}{SubISO}\\
\cline{3-6}
 & &  \#Embedding & Exec. time(ms) & \#Embedding & Exec. time(ms)\\			 				    			
      	 \hline
         \multirow{1}{*}{$Yq_{10}$/Yeast} & 85  & None & $6\times10^{5}$ & 1000 & 4\\  			 		
        \hline
        \multirow{1}{*}{$Yq_{15}$/Yeast} & 91  & None & $6\times10^{5}$ & 96 & 16\\	    					 \hline
         \multirow{1}{*}{$Yq_{20}$/Yeast} & 44  & None & $6\times10^{5}$ & 1000 & 31 \\  			 		
       	 \hline
         \multirow{1}{*}{$Yq_{25}$/Yeast} & 12  & None & $6\times10^{5}$ & 576 & 13\\  			 		
       	\hline
         \multirow{2}{*}{$Hq_{10}$/Human} & 13 & None & $6\times10^{5}$ & 1000 & 3\\  
         								& 44  & None & $6\times10^{5}$ & 396 & 19\\  				
       	\hline
         \multirow{2}{*}{$Hq_{15}$/Human}  & 24 & None & $6\times10^{5}$ & 24 & 66\\  	  			 		
      						    			& 62 & None & $6\times10^{5}$ & 1000 & 15830\\  	
       	\hline
         \multirow{2}{*}{$Hq_{20}$/Human} & 7  & None & $6\times10^{5}$ & 1000 & 27\\  			 	
       						    			& 11 & None & $6\times10^{5}$ & 1000 & 28\\  		   			 	 				    		
       	\hline
         \multirow{2}{*}{$Hq_{25}$/Human} & 31  & None & $6\times10^{5}$ & 1000 & 9\\  			 	
       						    			& 33  & None & $6\times10^{5}$ & 1000 & 44\\  		   			 	 				    		
       	\hline					    			
         \end{tabular} 
        }
  \end{center}
\end{table}

\begin{table}[!htb]
  \begin{center}
  \caption{Performance analysis of $\mathrm{SubISO}$ to deal with \texttt{QuickSI's} straggler queries identified from \texttt{CFL-Match}'s query sets}
 \scalebox{0.80}{
	 	\label{tab:Statistics of QuickSI Straggler queries}
	 		\begin{tabular}{|l|c|c|c|c|c|}	 
    \hline
    \multirow{2}{*}{Query set/Data graph} & \multirow{2}{*}{Q. graph id} & \multicolumn{2}{c|}{QuickSI} & %
    \multicolumn{2}{c|}{SubISO}\\
\cline{3-6}
 & &  \#Embedding & Exec. time(ms) & \#Embedding & Exec. time(ms)\\
       \hline
        \multirow{2}{*}{$CFLq_{15s}$/Human} & 74  & None & $6\times10^{5}$ & 1000 & 74\\    
       						    			& 78  & None & $6\times10^{5}$ & 1000 & 19\\
        \hline
         \multirow{2}{*}{$CFLq_{20s}$/Human} & 31  & None & $6\times10^{5}$ & 1000 & 45\\    
       						    			& 32 & None & $6\times10^{5}$ & 1000 & 16\\
       	\hline
         \multirow{2}{*}{$CFLq_{25s}$/Human} & 0 & None & $6\times10^{5}$ & 1000 & 38\\    
       						    			& 7  & None & $6\times10^{5}$ & 1000 & 16\\
       \hline
         \multirow{1}{*}{$CFLq_{10d}$/Human} & 76  & None & $6\times10^{5}$ & 0 & 31\\    
         									
       	\hline
         \multirow{2}{*}{$CFLq_{15d}$/Human} & 2  & None & $6\times10^{5}$ & 1000 & 5 \\    
       						    			& 57 & None & $6\times10^{5}$ & 755 & 6\\		
        \hline
         \multirow{2}{*}{$CFLq_{20d}$/Human} & 62  & None & $6\times10^{5}$ & 1000 & 7\\  
       						    			& 95  & None & $6\times10^{5}$ & 0 & 14\\    	       	
       	 \hline
         \multirow{2}{*}{$CFLq_{25d}$/Human} & 2  & None & $6\times10^{5}$ & 1000 & 95\\  			 	
       						    			& 55  & None & $6\times10^{5}$ & 1000 & 10\\
       	\hline    
        \end{tabular} 
        }
  \end{center}
\end{table}

\begin{table}[!htb]
  \begin{center}
  \caption{Performance analysis of $\mathrm{SubISO}$ to deal with \texttt{RI's} straggler queries identified from our generated query sets}
  \scalebox{0.80}{
	 	\label{tab:Statistics of RI Straggler queries1}
	 		\begin{tabular}{|l|c|c|c|c|c|}	 
    \hline
    \multirow{2}{*}{Query set/Data graph} & \multirow{2}{*}{Q. Graph id} & \multicolumn{2}{c|}{RI} & %
    \multicolumn{2}{c|}{SubISO}\\
\cline{3-6}
 & &  \#Embedding & Exec. time(ms) & \#Embedding & Exec. time(ms)\\			 				    			
        \hline
        \multirow{1}{*}{$Yq_{15}$/Yeast} & 62  & None & $6\times10^{5}$ & 1000 & 2\\	    					 \hline
         \multirow{2}{*}{$Yq_{20}$/Yeast} & 2  & None & $6\times10^{5}$ & 864 & 98 \\  			 		
       						    			& 13 & None & $6\times10^{5}$ & 1000 & 4 \\  		
       	 \hline
         \multirow{2}{*}{$Yq_{25}$/Yeast} & 37  & None & $6\times10^{5}$ & 1000 & 19\\  												  & 42  & None & $6\times10^{5}$ & 144 & 11\\  		 		
       	\hline
         \multirow{2}{*}{$Hq_{10}$/Human} & 7 & None & $6\times10^{5}$ & 1000 & 3\\  
         								& 44  & None & $6\times10^{5}$ & 396 & 19\\  				
       	\hline
         \multirow{2}{*}{$Hq_{15}$/Human}  & 2 & None & $6\times10^{5}$ & 1000 & 4\\  	  			 		
      						    			& 61 & None & $6\times10^{5}$ & 1000 & 4\\  	
       	\hline
         \multirow{2}{*}{$Hq_{20}$/Human} & 7  & None & $6\times10^{5}$ & 1000 & 27\\  			 	
       						    			& 11 & None & $6\times10^{5}$ & 1000 & 28\\  		   			 	 				    		
       	\hline
         \multirow{2}{*}{$Hq_{25}$/Human} & 31  & None & $6\times10^{5}$ & 1000 & 9\\  			 	
       						    			& 33  & None & $6\times10^{5}$ & 1000 & 44\\  		   			 	 				    		
       	\hline					    			
         \end{tabular} 
        }
  \end{center}
\end{table}

\begin{table}[!htb]
  \begin{center}
  \caption{Performance analysis of $\mathrm{SubISO}$ to deal with \texttt{RI's} straggler queries identified from the \texttt{CFL-Match}'s query sets}
 \scalebox{0.80}{
	 	\label{tab:Statistics of RI Straggler queries}
	 		\begin{tabular}{|l|c|c|c|c|c|}	 
    \hline
    \multirow{2}{*}{Query set/Data graph} & \multirow{2}{*}{Q. graph id} & \multicolumn{2}{c|}{RI} & %
    \multicolumn{2}{c|}{SubISO}\\
\cline{3-6}
 & &  \#Embedding & Exec. time(ms) & \#Embedding & Exec. time(ms)\\
       \hline
        \multirow{2}{*}{$CFLq_{10s}$/Human} & 44  & None & $6\times10^{5}$ & 1000 & 5\\    
       						    			& 62  & None & $6\times10^{5}$ & 1000 & 7\\
       	\hline					    
        \multirow{2}{*}{$CFLq_{15s}$/Human} & 70  & None & $6\times10^{5}$ & 1000 & 57\\    
       						    			& 81  & None & $6\times10^{5}$ & 954 & 79\\
        \hline
         \multirow{2}{*}{$CFLq_{20s}$/Human} & 8  & None & $6\times10^{5}$ & 1000 & 11\\    
       						    			& 20 & None & $6\times10^{5}$ & 1000 & 23\\
       	\hline
         \multirow{2}{*}{$CFLq_{25s}$/Human} & 13 & None & $6\times10^{5}$ & 1000 & 263\\    
       						    			& 24  & None & $6\times10^{5}$ & 969 & 16\\
       \hline
         \multirow{2}{*}{$CFLq_{10d}$/Human} & 10  & None & $6\times10^{5}$ & 1000 & 9\\   
         									& 23  & None & $6\times10^{5}$ & 1000 & 3\\
       	\hline
         \multirow{2}{*}{$CFLq_{15d}$/Human} & 39  & None & $6\times10^{5}$ & 1000 & 9\\
       						    			& 41  & None & $6\times10^{5}$ & 760 & 9\\       
       						    			 \hline
         \multirow{2}{*}{$CFLq_{20d}$/Human} & 47  & None & $6\times10^{5}$ & 1000 & 5\\
       						    			& 55  & None & $6\times10^{5}$ & 908 & 5\\ 	       	
       	 \hline
         \multirow{2}{*}{$CFLq_{25d}$/Human} & 65  & None & $6\times10^{5}$ & 477 & 197\\
       						    			& 96  & None & $6\times10^{5}$ & 1000 & 14\\
       	\hline    
        \end{tabular} 
        }
  \end{center}
\end{table}

\begin{table}[!htb]
  \begin{center}
  \caption{Performance analysis of $\mathrm{SubISO}$ to deal with \texttt{Glasgow's} straggler queries identified from our generated query sets}
  \scalebox{0.80}{
	 	\label{tab:Statistics of Glasgow Straggler queries1}
	 		\begin{tabular}{|l|c|c|c|c|c|}	 
    \hline
    \multirow{2}{*}{Query set/Data graph} & \multirow{2}{*}{Q. Graph id} & \multicolumn{2}{c|}{Glasgow} & %
    \multicolumn{2}{c|}{SubISO}\\
\cline{3-6}
 & &  \#Embedding & Exec. time(ms) & \#Embedding & Exec. time(ms)\\			 				    			
       			 \hline
         \multirow{1}{*}{$Yq_{20}$/Yeast} & 35  & None & $6\times10^{5}$ & 4 &  3410 \\  			 		
       	\hline
         \multirow{1}{*}{$Hq_{20}$/Human} & 18  & None & $6\times10^{5}$ & 96 & 1261\\  			 	
       			   			 	 				    		
       	\hline
         \multirow{2}{*}{$Hq_{25}$/Human} & 66  & None & $6\times10^{5}$ & 1000 & 55\\  			 	
       						    			& 79  & None & $6\times10^{5}$ & 1000 & 57\\  		   			 	 				    		
       	\hline					    			
         \end{tabular} 
        }
  \end{center}
\end{table}

\begin{table}[!htb]
  \begin{center}
  \caption{Performance analysis of $\mathrm{SubISO}$ to deal with \texttt{Glasgow's} straggler queries identified from the \texttt{CFL-Match}'s query sets}
 \scalebox{0.80}{
	 	\label{tab:Statistics of Glasgow Straggler queries}
	 		\begin{tabular}{|l|c|c|c|c|c|}	 
    \hline
    \multirow{2}{*}{Query set/Data graph} & \multirow{2}{*}{Q. graph id} & \multicolumn{2}{c|}{Glasgow} & %
    \multicolumn{2}{c|}{SubISO}\\
\cline{3-6}
 & &  \#Embedding & Exec. time(ms) & \#Embedding & Exec. time(ms)\\
       \hline
         \multirow{2}{*}{$CFLq_{25s}$/Human} & 78 & None & $6\times10^{5}$ & 0 & 276\\    
       						    			& 89  & None & $6\times10^{5}$ & 0 & 79\\     
       	\hline
         \multirow{1}{*}{$CFLq_{20d}$/Human} & 70  & None & $6\times10^{5}$ & 0 & 500\\
       						    			     	
       	 \hline
         \multirow{2}{*}{$CFLq_{25d}$/Human} & 10  & None & $6\times10^{5}$ & 0 & 4\\
       						    			& 58  & None & $6\times10^{5}$ & 0 & 322\\
       	\hline    
        \end{tabular} 
        }
  \end{center}
\end{table}

We have performed similar performance analysis of $\mathrm{SubISO}$ to deal with \texttt{QuickSI}'s straggler queries identified from our generated query sets and CFL-Match's query sets. Tables  \ref{tab:Statistics of QuickSI Straggler queries1} and\ref{tab:Statistics of QuickSI Straggler queries} present the performance analysis results of $\mathrm{SubISO}$ in terms of \texttt{number of embeddings} and \texttt{execution time} to deal with \texttt{QuickSI's} straggler queries identified from our generated query sets (table \ref{tab:StragglerqueriesofQuickSI}) and CFL-Match's query sets (table \ref{tab:CFLStragglerqueriesofQuickSI}), respectively. It can be observed from these tables that $\mathrm{SubISO}$ is able to find embeddings for each \texttt{QuickSI}'s straggler queries in a very short execution time.

Finally, we have performed a performance analysis of $\mathrm{SubISO}$ to deal with \texttt{RI}'s straggler queries identified from our generated query sets and CFL-Match's query sets. Tables  \ref{tab:Statistics of RI Straggler queries1} and \ref{tab:Statistics of RI Straggler queries} present the performance analysis results of $\mathrm{SubISO}$ in terms of \texttt{number of embeddings} and \texttt{execution time} to deal with \texttt{RI's} straggler queries identified from our generated query sets (table \ref{tab:StragglerqueriesofRI}) and CFL-Match's query sets (table \ref{tab:CFLStragglerqueriesofRI}), respectively. Again, similar observations can be made from these tables that $\mathrm{SubISO}$ is able to find embeddings for each \texttt{RI}'s straggler queries in a very short execution time.

\section{Disscussion}\label{sec:Disscussion}
In this section, we present a discussion on the impact of limiting the number of recursive calls, followed by a proof of the \textit{soundness} and \textit{completeness} of our proposed $\mathrm{SubISO}$ method.
\begin{figure*}[!htb]
  \centering
  \subfigure[Yeast]{\includegraphics[scale=0.23,keepaspectratio]{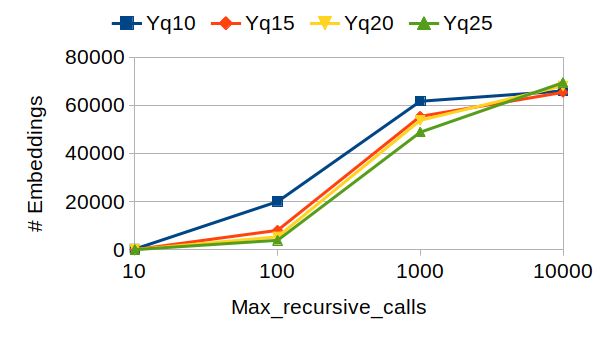}}\quad
  \subfigure[Human]{\includegraphics[scale=0.23,keepaspectratio]{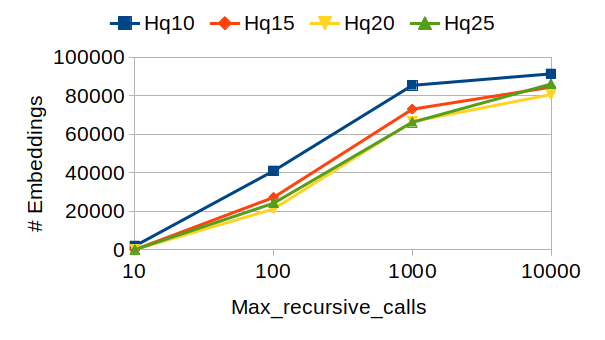}}\quad
   \subfigure[Hprd]{\includegraphics[scale=0.23,keepaspectratio]{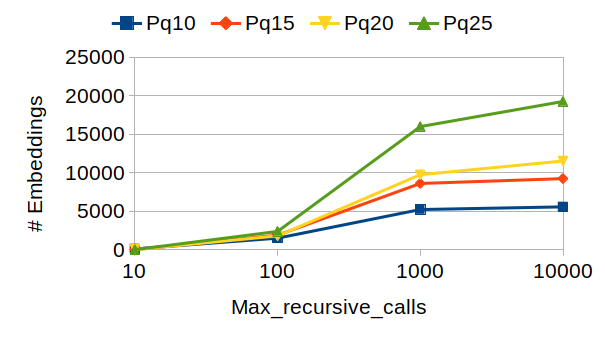}}
  \caption{Maximum recursive calls vs number of embeddings on our generated query sets}
  \label{fig:NumberofEmbeddingsVSMaxreurssivecalls}
\end{figure*}

\begin{figure*}[!htb]
  \centering
  \subfigure[Yeast]{\includegraphics[scale=0.23,keepaspectratio]{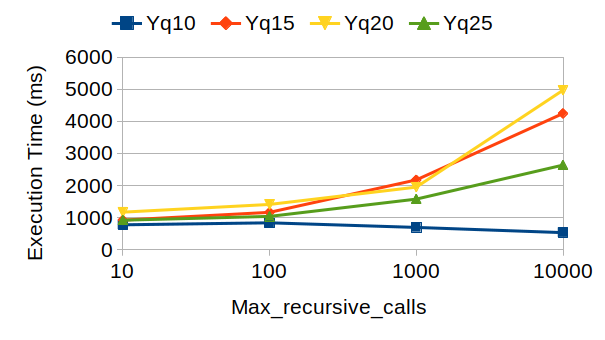}}\quad
  \subfigure[Human]{\includegraphics[scale=0.23,keepaspectratio]{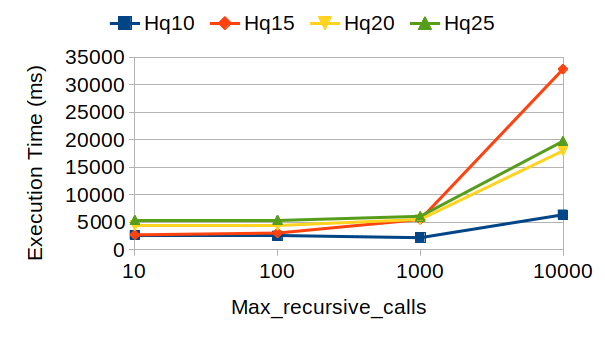}}\quad
   \subfigure[Hprd]{\includegraphics[scale=0.23,keepaspectratio]{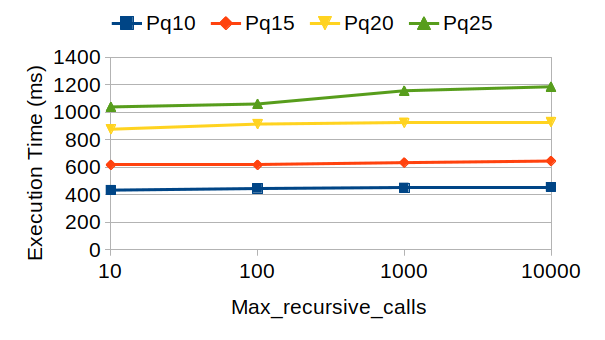}}
  \caption{Maximum recursive calls vs execution time on our generated query sets}
  \label{fig:ExecutiontimeVSMaxreurssivecalls}
\end{figure*}
  
\subsection{Impact of limiting the number of recursive calls} \label{subsec:WhyweLimitMaximumRecursiveCall}
\begin{figure}[!htb]
  \centering
  \subfigure[Dense query graph]{\includegraphics[scale=0.23,keepaspectratio]{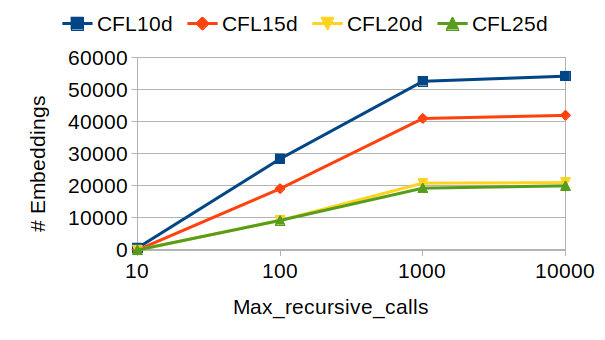}}\quad
  \subfigure[Sparse query graph]{\includegraphics[scale=0.23,keepaspectratio]{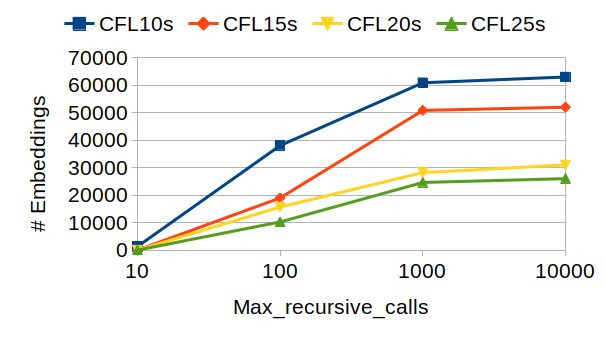}}
  \caption{Maximum recursive calls vs number of embeddings on \texttt{CFL-Match}'s query sets}
  \label{fig:CFLNumberofEmbeddingsVSMaxreurssivecalls}
\end{figure}

\begin{figure}[!htb]
  \centering
  \subfigure[Dense query graph]{\includegraphics[scale=0.23,keepaspectratio]{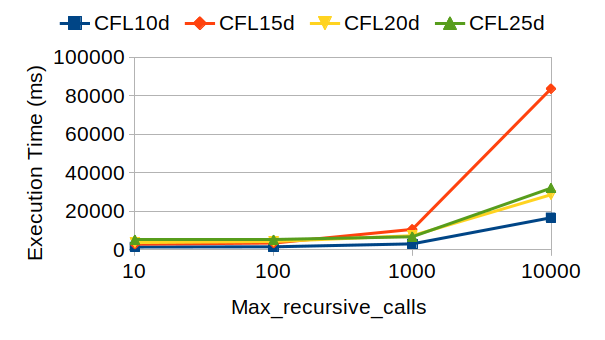}}\quad
  \subfigure[Sparse query graph]{\includegraphics[scale=0.23,keepaspectratio]{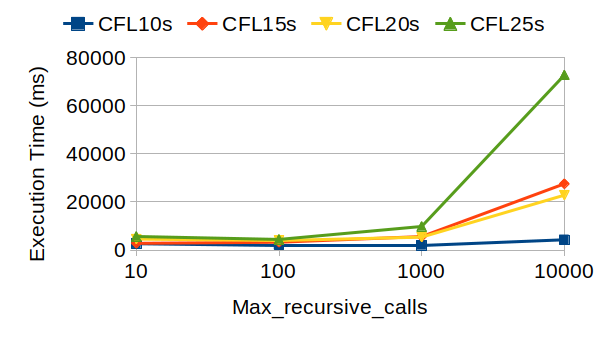}}
  \caption{Maximum recursive calls vs execution time on \texttt{CFL-Match}'s query sets}
  \label{fig:CFLExecutiontimeVSMaxreurssivecalls}
\end{figure}

It can be observed from the evaluation results presented in the previous section that limiting recursive calls $\mathrm{SubISO}$ play an important role, mainly to deal with straggler queries of the existing state-of-the-art methods. Figure \ref{fig:NumberofEmbeddingsVSMaxreurssivecalls} presents a visualization of the impact of increasing the number of recursive calls ($\eta$) over the number of identified embeddings for our generated query sets. It can be observed from this figure that the number of embeddings increases rapidly when we increase the value of $\eta$ to 1000. However, when the value of $\eta$ goes beyond 1000, the number of embeddings shows a slow increase. On the other hand, figure \ref{fig:ExecutiontimeVSMaxreurssivecalls} presents a visualization of the impact of increasing the number of recursive calls over the execution time of $\mathrm{SubISO}$. It can be observed from this figure that initially execution time increases linearly; however, it increases exponentially, except the case of Hprd dataset, when the value of $\eta$ exceeds 1000. Similar trends can be observed from figures \ref{fig:CFLNumberofEmbeddingsVSMaxreurssivecalls} and \ref{fig:CFLExecutiontimeVSMaxreurssivecalls} which visualizes the impact of increasing the number of recursive calls over the number of identified embeddings and execution time, respectively in case of using CFL-Match's query sets.     

Based on the above observations, it can concluded that restricting the number of recursive calls ($\eta$), saves a large amount of execution time and obtains a good number of embeddings. However, finding an ideal value of $\eta$ to balance between the execution time and number of embeddings remains an empirical analysis problem. 


%

On analysis, we found that the distribution of query graph embeddings is not uniform across different candidate regions. Some candidate regions contain a larger number of embeddings, whereas other candidate regions have fewer number of embeddings. Moreover, sometimes the generic subgraph search function (i.e., \texttt{SubgraphSearch()}) gets stuck in a candidate region when it is not able to obtain correct match for some set of query graph vertices, resulting in the utilization of the excessive number of recursive calls.  Due to this, many algorithms show exponential behavior for some set of query/data graphs. In order to deal with this problem, we have considered restricting the number of recursive calls of the \texttt{SubgraphSearch} function for each candidate region so that it could switch to explore another candidate region after a certain number of attempts to find a solution (embedding) in the region under consideration. However, limiting recursive calls of the \texttt{SubgraphSearch} function saves time, but at the cost of possibly missing some solutions that can be acceptable because the number of solutions for subgraph isomorphism problem is generally very large. Moreover, limiting the number of recursive calls makes $\mathrm{SubISO}$ capable to deal with straggler queries on which state-of-the-art methods show exponential behavior.

\subsection{\texorpdfstring{$\mathrm{SubISO}$}is sound and complete} \label{subsec:SubISOisSoundandComplete}
In this section, we prove that our proposed $\mathrm{SubISO}$ algorithm is \textit{sound} and \textit{complete}. By \textit{sound} we mean $\mathrm{SubISO}$ finds only those embeddings that follow the definition of subgraph isomorphism given in definition \ref{def:SubgraphIsomorphism}. On the other hand, by \textit{complete} we mean $\mathrm{SubISO}$ can enumerate all instances of a query graph that are embedded in the data graph. These properties are formally stated and proved in the following paragraphs:

\begin{theorem}
The proposed $\mathrm{SubISO}$ algorithm is sound and complete.
\end{theorem}

\begin{proof}
Let $\mathpzc{M_1}$ be the set of all embeddings of a query graph $G_{q}$ in a data graph $G_{d}$ obtained by $\mathrm{SubISO}$. Let $\mathpzc{M_2}$ be the set of all embeddings of $G_{q}$ in $G_{d}$ that follow definition \ref{def:SubgraphIsomorphism}.\\
In order to prove that $\mathrm{SubISO}$ is sound, it is sufficient to prove that $\mathpzc{M_1} \subseteq \mathpzc{M_2}$

Let $m \in \mathpzc{M_1}$. Therefore, $m$ is the set of all ordered pairs of type $<$\!$u,f(u)$\!$>, \forall u \in V(G_q)$ and $f:V(G_q) \rightarrow V(G_d)$ is a  mapping that follows $\mathrm{SubISO}$.\\

As per definition \ref{def:SubgraphIsomorphism}, we have to first prove that $f:V(G_q) \rightarrow V(G_d)$ is an injective mapping.\\

Let if $u_1$ and $u_2$ be any two different vertices of $V(G_q)$. Since step 6 of the algorithm \ref{Algo:SubgraphSearch} always selects unmatched element of $\psi_\mathcal{R}(u_1)$ and $\psi_\mathcal{R}(u_2)$ for matching purpose, when an element from $\psi_\mathcal{R}(u_1)$ is selected as a match for $u_1$ it is marked as matched and it can never be selected as a match for any other vertex of $G_q$. Hence, for any two different vertices of $V(G_q)$, their corresponding images $f(u_1)$ and $f(u_2)$ can not be same. This proves that $f$ is injective. 
Now, since the label of vertices is one of the isomorphic invariants that $f$ follows, the label of a vertex $u$ and its image $f(u)$ must be same. This proves the first condition of definition \ref{def:SubgraphIsomorphism}.\\

Algorithm \ref{Algo:IsJoinable} ensures that if $u_1$ and $u_2$ are connected through an edge then there must exist an edge between $f(u_1)$ and $f(u_2)$. This proves the second condition of definition \ref{def:SubgraphIsomorphism}. Hence, $f$ is a mapping that follows definition \ref{def:SubgraphIsomorphism}, and it can be concluded $m \in \mathpzc{M_2}$, and accordingly $\mathpzc{M_1} \subseteq \mathpzc{M_2}$.\\
Thus, $\mathrm{SubISO}$ is sound.

Now, in order to prove that $\mathrm{SubISO}$ is complete, it is sufficient to show that $\mathpzc{M_1} = \mathpzc{M_2}$

Since we have already proved that $\mathpzc{M_1} \subseteq \mathpzc{M_2}$, we need to prove only $\mathpzc{M_2} \subseteq \mathpzc{M_1}$.\\ 
Let $m \in \mathpzc{M_2}$. Therefore, $m = \{<$\!$u,f(u)$\!$>\mid~f$ is a subgraph isomorphic map from $G_{q}$ to $G_{d}$ and $u \in G_q$\}.\\
Let $\hat{u}$ be a pivot vertex of $G_{q}$ selected by $\mathrm{SubISO}$ that always exist if we follow the steps of algorithm \ref{Algo:PivotVertexSelection}.\\
Since  $\hat{u} \in G_{q}$  and $f$ is a subgraph isomorphic map from $G_{q}$ to $G_{d}$, 
$<$\!$\hat{u}, f(\hat{u})$\!$> \in m $.\\
Since all possible subgraph isomorphic images of $\hat{u}$ is in $\psi(\hat{u})$, $\Rightarrow f(\hat{u}) \in \psi(\hat{u})$\\
Let $\mathcal{R}$ be the candidate region corresponding to $ f(\hat{u})$ that always exist if we follow the steps of algorithm \ref{Algo:RegionExploration}.\\
Since $\mathrm{SubISO}$ explores candidate region corresponding to every element of $\psi(\hat{u})$ and finds all possible embeddings in that region using algorithm \ref{Algo:SubgraphSearch}, it also explores the candidate region $\mathcal{R}$ and finds all embeddings in $\mathcal{R}$ out of which at least one embedding will coincide with $m$.\\
Hence, $m \in \mathpzc{M_1}$, and accordingly $\mathpzc{M_2} \subseteq \mathpzc{M_1}$.\\
This proves that $\mathrm{SubISO}$ is complete.
\end{proof}

\section{Conclusion and Future Work}\label{sec:ConclusionAndFutureWork}
In this paper, we have proposed a new objective function, which is based on the notion of eccentricity and some isomorphic invariants of the query graph vertices. The objective function aims to minimize both number and size of the candidate regions in the data graph for efficiently enumerating all subgraph isomorphisms of the query graph. Using the objective function to locate pivot vertex of the query graph, we have designed and implemented $\mathrm{SubISO}$ algorithm, which consists of $OptimalPivotVertexSelection(),$ $RegionExploration()$, and $SubgraphSearch()$ functions to find solutions of the subgraph isomorphism problem. We have compared the performance of $\mathrm{SubISO}$ with two popular state-of-art subgraph isomorphism methods, $\mathrm{Turbo_{ISO}}$ and \texttt{QuickSI}, over three benchmark datasets -- \texttt{Human}, \texttt{Yeast}, and \texttt{Hprd}. We have also introduced a new parameter $\eta$ to restrict the maximum recursive calls of the $SubgraphSearch()$ function and studied its impact on the execution time and the number of identified embeddings of the query graph. On empirical analysis, we found that restricting the maximum recursive calls makes $\mathrm{SubISO}$ able to deal with the straggler queries for which state-of-the-art methods show exponential behavior. The program codes of $\mathrm{SubISO}$ and our generated query sets for all benchmark datasets will be made publicly available after acceptance and publication of this research paper.    

Some of the future directions of research lie in (i) understanding and characterizing the spread of query graph embeddings in different candidate regions of the data graph, (ii) estimating a suitable value of the $\eta$ parameter based on the characteristics of the query and/or data graphs, (iii) exploring parallel computing platforms for executing region exploration and subgraph matching processes together, and (iv) characterizing and identifying straggler queries for a given subgraph isomorphism algorithm and data graph without actually executing the program codes.


\bibliographystyle{abbrvnat}
\bibliography{Reference}

\end{document}